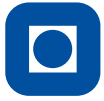

NTNU – Trondheim
Norwegian University of
Science and Technology

**Katina Kralevska**

# Applied Erasure Coding in Networks and Distributed Storage

Thesis for the degree of Philosophiae Doctor

Trondheim,   March 2018

Department of Telematics,
Faculty of Information Technology, Mathematics and Electrical Engineering,
NTNU, Norwegian University of Science and Technology
Trondheim, Norway

# Abstract

The amount of digital data is rapidly growing. There is an increasing use of a wide range of computer systems, from mobile devices to large-scale data centers, and important for reliable operation of all computer systems is mitigating the occurrence and the impact of errors in digital data.

The demand for new ultra-fast and highly reliable coding techniques for data at rest and for data in transit is a major research challenge. Reliability is one of the most important design requirements. The simplest way of providing a degree of reliability is by using data replication techniques. However, replication is highly inefficient in terms of capacity utilization. Erasure coding has therefore become a viable alternative to replication since it provides the same level of reliability as replication with significantly less storage overhead.

The present thesis investigates efficient constructions of erasure codes for different applications. Methods from both coding and information theory have been applied to network coding, Optical Packet Switching (OPS) networks and distributed storage systems. The following four issues are addressed:

– Construction of binary and non-binary erasure codes;

– Reduction of the header overhead due to the encoding coefficients in network coding;

– Construction and implementation of new erasure codes for large-scale distributed storage systems that provide savings in the storage and network resources compared to state-of-the-art codes; and

– Provision of a unified view on Quality of Service (QoS) in OPS networks when erasure codes are used, with the focus on Packet Loss Rate (PLR), survivability and secrecy.

A major part of the present thesis is the study of both theoretical and practical aspects of code constructions for distributed storage systems. Distributed storage systems typically employ commodity hardware, often mounted in racks, so that the system can be scaled at a low cost. The components may suffer from failures and other factors, such as software glitches and machine reboots during maintenance operations, that result in unavailability of the stored data. The reliability provided by 3-replication is an accepted industry standard for incorporating reliability into storage systems. Nevertheless, the relentless data growth has made erasure coding a valuable alternative to 3-replication, and hence many distributed storage systems such as Hadoop Distributed File System (HDFS), OpenStack SWIFT and Microsoft Azure employ Reed-Solomon (erasure) codes.

New metrics for efficient erasure coding solutions have been identified in the literature. Some of these metrics, that are also studied in the present thesis, include: 1) reliability, 2) storage efficiency, 3) repair bandwidth, 4) disk-I/O, 5) repair locality, and 6) update complexity. Each of these metrics has a different relevance to a specific system depending on the system's architecture and the workload.

In the present thesis, we propose two novel constructions of erasure codes for distributed storage. The first construction is called *HashTag Erasure Codes (HTECs)*. HTECs are storage-reliability optimal meaning that they offer maximum fault tolerance for the consumed storage. HTECs are the first codes in the literature that reduce the repair bandwidth for both single and multiple failures for an arbitrary sub-packetization level. The bandwidth savings can go up to 70% and 30% compared to RS codes for single and double failures, respectively. HTECs address also the practical problem of disk I/O operations with the focus on reducing the number of random operations that access locations on the storage devices in a non-contiguous manner. The second construction of erasure codes belongs to the class of Locally Repairable Codes (LRCs). The proposed *Balanced Locally Repairable Codes (BLRCs)* are suitable for applications that require a low repair locality for single and double failures, low storage overhead, high reliability and low update complexity.

The present thesis therefore provides new code constructions and demonstrates how these codes are applied to network coding, OPS networks and distributed storage systems.

# Preface

This dissertation is submitted in partial fulfillment of the requirements for the degree Philosophiae Doctor (PhD) at NTNU, Norwegian University of Science and Technology. The presented work was carried out at the Department of Telematics (ITEM) in the period October 2012 – August 2016 under the supervision of Associate Professor Harald Øverby and the co-supervision of Professor Danilo Gligoroski and Assistant Professor Gergely Biczók.

# Acknowledgement

With deep sense of gratitude, I thank my supervisor Assoc. Prof. Harald Øverby for his guidance and support throughout the PhD period. He has always had an open door and time for me. I would like to thank Prof. Danilo Gligoroski for his constant support and fruitful discussions. The collaboration with him was both inspiring and encouraging. Many thanks to Assis. Prof. Gergely Biczók for his useful advice and insights. I would further like to thank Prof. Steinar Hidle Andresen for accepting me as an IAESTE trainee at ITEM and later supporting me to pursue a PhD.

During my PhD period I have cooperated with the Transfer Technology Office (TTO) at NTNU. I am sincerely grateful to Torbjørn Rostad and the rest of the TTO team who saw a market potential in my research. Per Simonsen, Rune E. Jensen, Sindre B. Stene and Kjetil Babington have contributed greatly in the establishment of the startup company MemoScale by creating a market value and applying the research in industrial products. I would like to thank my co-authors Zoran Hadzi-Velkov, Yanling Chen, Tewelde D. Assefa and Yuming Jiang for the collaboration. Many thanks to my office-mate Chengchen Hu for the useful discussions. I would like to thank Gianfranco Nencioni, Poul E. Heegaard, Katrien De Moor, Mona Nordaune, Randi Flønes and the rest of my colleagues who have made ITEM such an enjoyable place to work at. Big thanks to my friends in Trondheim who have always been a great company for going out and exploring new places. Special thanks to Elissar Khloussy for being such a wonderful friend. My dear friends Atina and Jasmina from Macedonia have always been good friends even though we live far away from each other.

My final thanks are reserved to my dearest ones. I would like to thank my parents and brother for supporting me in all my decisions. Special thanks to my grandparents for their constant care. Last but not least, I would like to thank Henry for his understanding and happy moments.

# Contents

# List of Figures

# List of Tables

# List of Acronyms

**AMDS** Almost-Maximum Distance Separable.

**ARQ** Automatic Repeat reQuest.

**BCH** Bose-Chaudhuri-Hocquenghem.

**BHP** Burst Header Packet.

**CPT** Coded Packet Transport.

**CRC** Cyclic Redundancy Checksum.

**DBR** Data Burst Redirection.

**DWDM** Dense Wavelength Division Multiplexing.

**ECC** Error-Correcting Code.

**FDL** Fiber Delay Line.

**FEC** Forward Error Correction.

**HARQ** Hybrid Automatic Repeat reQuest.

**HDFS** Hadoop Distributed File System.

**HTECs** HashTag Erasure Codes.

**LDGM** Low-Density Generator Matrix.

**LDPC** Low Density Parity Check.

**LNC** Linear Network Coding.

**LT** Luby Transform.

**MBR** Minimum-Bandwidth Regenerating.

**MDS** Maximum Distance Separable.

**MRD** Maximum Rank Distance.

**MSCR** Minimum Storage Collaborative Regenerating.

**MSR** Minimum-Storage Regenerating.

**MTFF** Mean Time to First Failure.

**MTTDL** Mean Time To Data Loss.

**NLPRS** Network Layer Packet Redundancy Scheme.

**NMDS** Near-Maximum Distance Separable.

**Non-MDS** Non-Maximum Distance Separable.

**OBS** Optical Burst Switching.

**OPS** Optical Packet Switching.

**PLR** Packet Loss Rate.

**PM-MSR** Product-Matrix-MSR.

**PRNG** Pseudo-Random Number Generator.

**QoS** Quality of Service.

**RAID** Redundant Arrays of Inexpensive Disks.

**RAM** Random-Access Memory.

**RLNC** Random Linear Network Coding.

**RS** Reed Solomon.

**RSA** Rivest-Shamir-Adleman.

**RTT** Round Trip Time.

**SECDED** Single-Error-Correcting and Double-Error-Detecting.

**SHEC** Shingled erasure codes.

**SSAC** Small Set of Allowed Coefficients.

**WRON** Wavelength Routed Optical Networks.

# Part I

# Summary of the Thesis

# Chapter 1
# Introduction

## 1.1 Thesis Structure

The present thesis is a collection of papers which is in accordance with NTNU rules for PhD studies. It is divided into two main parts:

– **Part I: Summary of the Thesis**

– **Part II: Included Papers**

**Part I** is a comprehensive summary of the present thesis. It consists of three chapters:

– The *Introduction* chapter (Chapter 1) presents the motivation for applying erasure codes in different networks.

– The *Background and Related Works* chapter (Chapter 2) gives the necessary background to understand the contributions of the thesis. It also reviews the state-of-the-art for erasure coding, algorithms for header compression in network coding, code constructions for distributed storage and QoS metrics in OPS networks. Some of the challenges faced by the coding and the networking communities that are addressed in the present thesis are listed in the end of each section.

– The *Contributions and Concluding Remarks* chapter (Chapter 3) presents the research questions answered in the present thesis and the research contributions and results obtained during the PhD period. It also puts the presented research results into a wider context by comparing them to selected references. Finally, conclusions followed by suggestions for future work are presented.

**Part II** consists of 7 papers where 6 are published and 1 is submitted for publication (Table 3.1).

## 1.2   Motivation

The relentless data growth brings enormous challenges, as well as incredible research and business opportunities. IDC [IDC12] estimates that the total amount of digital data created, replicated and consumed will reach 40000 exabytes, or 40 trillion gigabytes, by 2020. From now until 2020, the amount of digital data is expected to double every two years, as shown in Figure 1.1.

Reliable communication through unreliable media is paramount in modern communication systems. Reliable communication requires that all intended receivers of the data receive the data intact, i.e. data must be transferred without errors or loss. The demand for efficient and reliable communication has been accelerated even more by the emergence of large-scale, high-speed data networks for exchange, processing and storage of digital information in public and private spheres. Reliability is achieved by adding redundancy at different levels in the protocol stack. One way is to use Forward Error Correction (FEC) codes as error correction or erasure codes. FEC codes preprocess the data in such a way to provide recovery after data corruption [CC81].

Erasure coding has emerged as a compliment, or an alternative, to Automatic Repeat reQuest (ARQ) and replication in communication networks and distributed storage systems, respectively.

Using erasure coding, or combining it with ARQ, is a better solution instead of only using ARQ. When the PLR is very high, then retransmissions happen frequently and the system throughput is reduced significantly. In this case, combining ARQ with FEC known as Hybrid Automatic Repeat reQuest (HARQ) [CC84, LY82] is useful. ARQ is not feasible for unidirectional broadcast networks or real-time applications, because a return channel may not exist or the Round Trip Time (RTT) delay may be too large. Additionally, when the number of users is very large, scalability issues may prevent the use of return channels. In all these scenarios, the use of erasure codes is imminent.

Replication is not an appropriate enabler of the exabyte era because it increases enormously the storage overhead. Let us take the following example where 3-replication and an $(9,6)$ erasure code provide a similar level of reliability. Three copies of the same file are stored with 3-replication, while the file is divided into 6 fragments and 3 redundancy fragments when constructed with an $(9,6)$ erasure code. The storage overhead is 200% with 3-replication, while it is only 50% with the erasure code. In this example, erasure coding reduces the cost of storage by 150%, which is a tremendous cost saving when storing exabytes of data. Erasure coding has a clear advantage over replication as it provides the same level of reliability with less storage overhead [WK02].

Although many different erasure codes have been developed, there is no erasure code construction that provides simultaneously optimal performance and reliability. Maximum Distance Separable (MDS) codes such as Reed-Solomon codes [RS60] are fault tolerant optimal, but they are computationally expensive in practice. Low Density Parity Check (LDPC) codes [Gal63] do not offer the same fault tolerance as MDS codes, but tend to be computationally inexpensive and may have regular or irregular fault tolerance. Random Linear Network Coding (RLNC) is a flexible coding scheme that does not follow a predesigned code. Depending on the finite field size, the codes can be fault tolerant optimal. Then the codes are computationally demanding as Reed-Solomon codes.

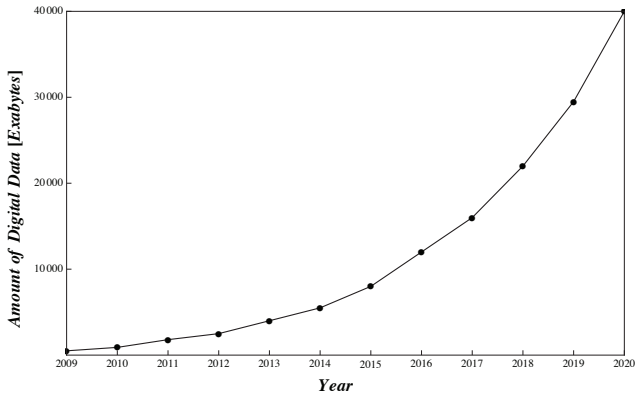

**Figure 1.1:** 50-fold growth of the amount of digital data from 2010 to 2020 [IDC12].

Both the underlying system and the application have a huge impact on which coding scheme gives the best performance. For instance, erasure coding has been currently deployed in Windows Azure [HSX$^+$12], big data analytics clusters (e.g. Facebook Analytics Hadoop cluster [SAP$^+$13]), archival storage systems, and peer-to-peer storage systems like Cleversafe. Facebook has recently reported that it is archiving old data using a classical Reed-Solomon erasure code implemented on the top of HDFS[SAP$^+$13], while Microsoft uses a pyramid code as the main storage primitive of its Azure storage service [HCL13]. Reed-Solomon codes, that are the essential building blocks in RAID 6, are optimal in terms of storage capacity utilization in large-scale distributed storage systems, but they perform poorly in terms of other system resources such as disk I/O and network bandwidth. During a recovery of lost or otherwise unavailable data, classical Reed-Solomon codes require a large amount of data to be read and transferred across the network.

Accordingly, the work presented in this thesis has been directed towards constructing efficient erasure codes for different applications. The four topics covered in the present thesis are:

– Construction of binary erasure codes (Paper 1 and Paper 2);

– Reduction of the header overhead due to coding coefficients in network coding (Paper 3);

- Construction of erasure codes for practical use in distributed storage systems (Paper 4, Paper 5 and Paper 6); and

- Application of erasure codes to increase the QoS in OPS/OBS networks (Paper 7).

# Chapter 2
# Background and Related Works

In this Chapter we review the background and the related works in the main research areas of this dissertation: erasure coding (Section 2.1), header compression algorithms for network coding (Section 2.2), code constructions for distributed storage systems (Section 2.3) and optical packet switched networks (Section 2.4).

## 2.1 Coding Theory

Data reliability, regardless of the medium, is achieved by adding redundancy. FEC (Forward Error Correction) codes are either used as *error correction* or *erasure codes*. Error correction codes are usually used at lower layers of the protocol stack, either as standalone codes or in conjunction with error detection checksums (e.g. Cyclic Redundancy Checksum (CRC))[PB61]. The upper layers deal with erasures (missing data units) after reception or storage/retrieval. Erasure codes are typically used in situations where the exact positions of the missing data units are known *a priori* (e.g. disk array). In many cases, error correction and erasure codes use the same encoding algorithm but have different decoding algorithms.

Since FEC codes treat each symbol as an element of a finite field and they perform extensively operations in finite fields, i.e. Galois fields, first a very brief introduction to Galois fields is given. For a more detailed description of Galois fields, please refer to [LN86].

### 2.1.1 Galois fields

Finite fields $\mathbf{F}_q$, also known as Galois fields $GF(q)$, are fundamental to coding theory. We use both notations $\mathbf{F}_q$ and $GF(q)$ interchangeably through the present thesis. The main advantage of using a Galois field is its closure property. A field is closed under both addition and multiplication meaning that the result of addition and multiplication of field elements is still a field element. Both operations are associative, distributive and commutative. Every non-zero element has a multiplicative inverse and every element has an inverse additive (negative) element. Working with data in transit or with data at rest in a Galois field means that the data elements are mapped into field elements, the performed operations follow the rules of the field and the data is reconstructed by inverse mapping.

The *order* of $GF(q)$ is the number of elements in the field. There exists a finite field of order $q$ if, and only if, $q$ is a prime power, i.e. $q = p^m$ where $p$ is a *prime number* called the *characteristic* of the field, and $m$ is a positive integer. If $m = 1$, then the field is called a *prime field*. Working in a prime field $GF(p)$ is quite simple. The prime field is the set of elements from 0 to $p-1$ under the operations of addition and multiplication modulo $p$.

If $m \geq 2$, then the field is called an *extension field*. Galois fields of order $2^m$ are called *binary extension fields*. These fields are used ubiquitously in coding theory and cryptography. One way of representing the elements in $GF(2^m)$ is to use a set of polynomials of degree at most $m-1$ where the coefficients are from the binary field $GF(2) = \{0, 1\}$:

$$GF(2^m) = \{c_{m-1}x^{m-1} + c_{m-2}x^{m-2} + \ldots + c_1 x^1 + c_0 x^0 : c_i \in \{0, 1\}\}. \quad (2.1)$$

Thus, the 4-bit element $a = (1, 1, 1, 0)_{2^4}$ has the following polynomial representation $a(x) = 1x^3 + 1x^2 + 1x + 0$.

The field $GF(2^m)$ is defined over an *irreducible polynomial* $f(x)$ of degree $m$ with coefficients from $GF(2)$. An irreducible polynomial is analogous to a prime number in that it cannot be factored as a product of polynomials each of degree less than $m$. Addition in $GF(2)$ is done with the bitwise XOR operator, while multiplication is performed with the bitwise AND operator. Addition in $GF(2^m)$ is the usual addition of polynomials with coefficient arithmetic performed modulo 2, while multiplication is more complex. Multiplication of two field elements $a$ and $b$ is performed by polynomial multiplication of $a(x)$ and $b(x)$ and then the product $a(x)b(x)$ is reduced modulo the irreducible polynomial $f(x)$.

Another useful property of any finite field $GF(q)$ is that the set of all nonzero elements $GF(q)^\times = GF(q) \setminus 0$ form a *multiplicative cyclic group* $(GF(q)^\times, \times)$. That means that any nonzero element $a \in GF(q)^\times$ can be represented as a power of a single element $\alpha \in GF(q)^\times$. Such a generator $\alpha$ is called a *primitive element* of the finite field. Powers of $\alpha$ repeat with a period length of $q-1$, thus, $\alpha^{q-1} = \alpha^0 = 1$. This makes multiplication of two elements $a = \alpha^i$ and $b = \alpha^j$ quite simple and fast. If we write $i = \log(a)$ and $a =$antilog$(i)$, then the product of $a, b \in GF(2^m)$ is computed as $ab =$antilog$(\log(a) + \log(b)) \mod 2^m - 1$.

### 2.1.2   Erasure Coding

Erasure codes, in particular linear block codes, are most appropriate for the applications such as data transmission and storage, which are of interest to the present thesis. The background presented in this subsection is essential to fully understand the work in the present thesis. Some terminology and definitions can be found in [LC83, Pla05].

A block encoder, according to certain rules, transforms each input message of $k$ information symbols into a message of $n$ symbols ($n > k$). This message of $n$ symbols is called a *codeword*. If the alphabet of the information source contains $q$ different digits, then there are a total of $q^k$ distinct messages of $k$ symbols. To each of the $q^k$ possible messages a unique codeword is assigned. This set of $q^k$ codewords of $n$ symbols is called a *q-ary block code of length n*.

**Definition 2.1.** A block code of length $n$ and $q^k$ codewords is called a *linear (n,k) code* if and only if its $q^k$ codewords form a $k$-dimensional subspace of the vector space of all $n$-tuples over the finite field $GF(q)$.

The *code rate R*, or code efficiency, is defined as $R = \frac{k}{n}$. The error detection and error correction capabilities of a $(n, k, d)$ code are defined by its metric, the minimum distance $d$. Before defining $d$, it is first necessary to define the Hamming weight of a codeword and the Hamming distance between two codewords.

**Definition 2.2.** The *Hamming weight* $w_H(c)$ of a codeword $c$ is the number of non-zero elements in $c$.

**Definition 2.3.** The *Hamming distance* $d_H(c_1, c_2)$ between two codewords $c_1$ and $c_2$, that have the same number of elements, is the number of elements in which these two codewords differ.

Consider the two codewords $c_1 = (0, 1, 0, 0, 1)$ and $c_2 = (1, 1, 0, 1, 1)$. The Hamming weights of $c_1$ and $c_2$ are $w_H(c_1) = 2$ and $w_H(c_2) = 4$, respectively, while $d_H(c_1, c_2) = 2$.

**Definition 2.4.** The *minimum distance* of a $(n, k)$ block code $C$, denoted by $d$, is the minimum among the Hamming distances between any two different codewords from $C$. A block code $C$ of length $n$, $q^k$ codewords and minimum distance $d$ is denoted by $(n, k, d)$ code $C$.

**Theorem 2.5.** *[Sin64] For a $(n, k, d)_q$ linear code we have*

$$d \leq n - k + 1 \quad \text{(Singleton bound)}. \tag{2.2}$$

*Codes for which the equality holds are known as Maximum Distance Separable (MDS) codes.*

When MDS codes are used as erasure codes, the receiver can recover the $k$ source symbols from any subset of $k$ received symbols out of the $n$ encoded symbols.

The *Singleton defect* of a $(n, k, d)$ code $C$, that is defined as $s(C) = n - k + 1 - d$, measures how far away is $C$ from being a MDS code. Based on the Singleton defect, the codes are divided in two classes:

1. Optimal, or very close to optimal, ones known as MDS [MS78], Almost-Maximum Distance Separable (AMDS) [dB96] and Near-Maximum Distance Separable (NMDS) [DL95].

2. Suboptimal or Non-Maximum Distance Separable (Non-MDS) codes [GLW10, Haf05, HCL13, LMS+97].

**Definition 2.6.** A $(n, k, d)$ code $C$ is
(i) *$t$-error detecting code* iff its minimum distance is at least $t + 1$;
(ii) *$t$-error correcting code* iff its minimum distance is at least $2t + 1$.

As mentioned earlier, a linear code constitutes a subspace and thus any codeword can be represented by a linear combination of the basis vectors of the subspace, i.e. by a linear combination of linearly independent vectors. The basis vectors can be written as rows of a $k \times n$ matrix.

**Definition 2.7.** Any matrix $\mathbf{G} \in \mathbf{F}_q^{k \times n}$ whose rows form a basis of $C$ is called a *generator matrix* for $C$, i.e.

$$C = \{\mathbf{xG} \in \mathbf{F}_q^n : \mathbf{x} \in \mathbf{F}_q^k\} = \{\mathbf{y} \in \mathbf{F}_q^n\}. \tag{2.3}$$

The injective map $\mathbf{F}_q^k \rightarrowtail \mathbf{F}_q^n$, $\mathbf{x} \rightarrowtail \mathbf{xG}$ encodes all the $q$-ary words of length $k$ to words of length $n$.

A generator matrix is called *systematic* if it is of the form

$$\mathbf{G} = \left[I_k | P\right], \tag{2.4}$$

where $I_k$ is an identity matrix of order $k$ and $P$ is a $k \times (n - k)$ parity submatrix.

The code can either be systematic or non-systematic. The generator matrix of a systematic code has the same form as the matrix represented in (2.4). That means that a systematic linear code does not transform the original $k$ data symbols, but it only generates extra $r$ redundant symbols. If the first $k$ rows in $\mathbf{G}$ do not contain an identity matrix, then the code is non-systematic. That is to say, all $n$ generated symbols linearly depend on all original $k$ symbols via $\mathbf{G}$. Systematic codes are less computationally demanding than non-systematic, since they do not require processing of the original data.

Given a generator matrix $\mathbf{G}$ of a linear code we can derive its parity-check matrix $\mathbf{H}$ (and vice versa).

**Definition 2.8.** The *parity-check matrix* $\mathbf{H}$ for a $(n, k, d)$ linear code with a generator matrix $\mathbf{G}$ (2.4) is given by

$$\mathbf{H} = \left[-P^T | I_{n-k}\right], \tag{2.5}$$

**Figure 2.1:** Encoding of the source data **x** with the generator matrix **G** of a $(8, 5)$ MDS code.

Since the rows of **H** are linearly independent, it generates a $(n, n-k, d')$ linear code called the *dual code* of the $(n, k, d)$ linear code generated with **G**.

Codes with generator matrices that are sparse and balanced minimize the maximal computation time of computing any code symbol. The problem of constructing balanced and sparse codes was studied in [DSDY13, HLH16].

**Theorem 2.9.** *[DSDY13] Suppose $1 \leq k \leq n$ and $q > \binom{n-1}{k-1}$. Then there always exits a $(n, k)_q$ MDS code that has a generator matrix **G** satisfying the following two conditions:*

- *Sparsest: each row of **G** has weight $n - k + 1$; and*

- *Balanced: column weights of **G** differ from each other by at most one.*

Let us explain the encoding and decoding of data with a $(8, 5)$ MDS code. As shown in Figure 2.1, $k = 5$ source data units represented as a row vector **x** are encoded into $n = 8$ data units (a row vector **y**) with the systematic generator matrix **G** of a $(8, 5)$ MDS code. The coefficients $c_{i,j}$, $i = 1, \ldots, 5$ and $j = 1, 2, 3$, are elements from $GF(q)$. The encoder in Figure 2.2 performs this multiplication. Next, the data is transmitted through unreliable medium to the receiver. Some of the data units may get corrupted (lost) during the transmission. Since the code is MDS, at most 3 data units can be lost. In this example, $x_1, x_4, r_2$ are lost. Namely, the decoder has to receive at least 5 data units in order to reconstruct the source data. Recovery of the source data is done by

$$\mathbf{y}' = \mathbf{x}\mathbf{G}' \rightarrow \mathbf{x} = \mathbf{y}'\mathbf{G}'^{-1}, \quad (2.6)$$

where **x** is the source data and **y**' is the subset of $k$ components available at the decoder. The matrix **G**' is the subset of columns from **G** corresponding to the components of **y**'. The source data **x** can be reconstructed only if **G**' is non-singular. This means, in general cases, any $k \times k$ submatrix extracted from **G** has to be invertible in order to recover from at most $n - k$ lost data units. In the presented example, the matrix **G**' is obtained by deleting the columns in **G** that correspond to the lost data units. Decoding is done in two steps: first $\mathbf{G}'$ is inverted and then the data is decoded by computing $\mathbf{x} = \mathbf{y}'\mathbf{G}'^{-1}$ as shown in Figure 2.3.

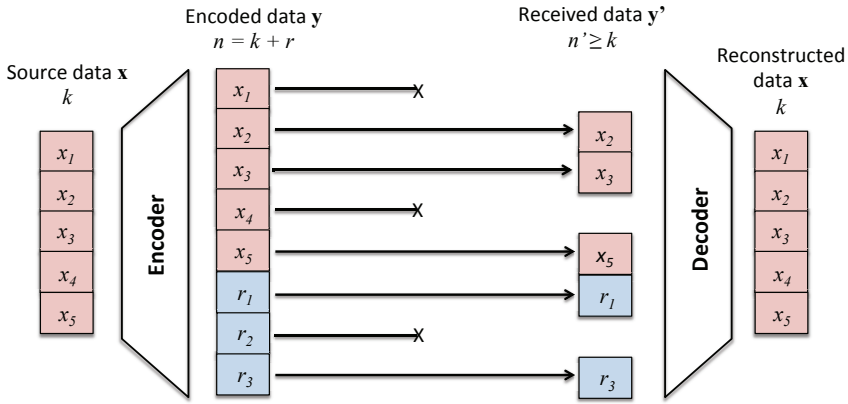

**Figure 2.2:** A graphical representation of the encoding/decoding process. The encoder encodes $k$ source units into $n$. The decoder has to receive at least $k$ data units in order to reconstruct the source data.

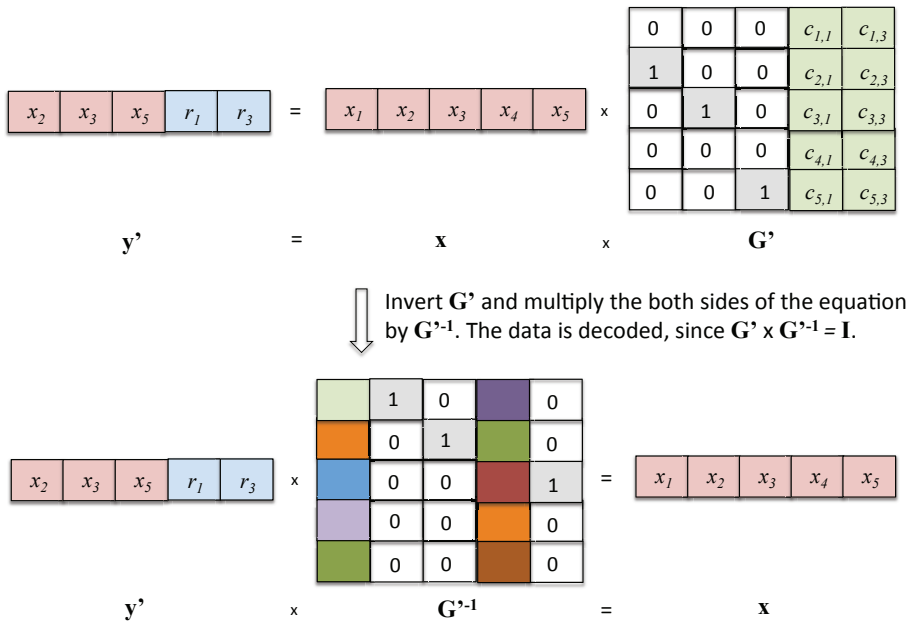

**Figure 2.3:** A graphical representation of the decoding process at the decoder.

There are numerous types of erasure codes. Some of the codes, such as Luby Transform (LT) [Lub02], Tornado [LMSS01a], and Raptor [Sho06] are protected by patents and hence their further development by third parties is problematic.

Among all codes in the class of block codes, cyclic codes are the most important from practical point of view. Bose-Chaudhuri-Hocquenghem (BCH) codes were discovered independently in the papers by Bose and Chaudhuri (1960) [BRC60] and Hocquenghem (1959) [Hoc59]. BCH codes are cyclic codes that have algebraic decoding algorithms.

Reed Solomon (RS) codes were first described in a paper by Reed and Solomon in 1960 [RS60]. They are non-binary BCH codes defined by the parameters: $n = q - 1, n - k = 2t, d = 2t + 1$. Since the minimum distance is $n - k + 1$, RS codes are MDS codes. RS codes are widely implemented in storage devices and communication standards. RS encoding is relatively straightforward, but decoding is complex despite the significant efficiency improvements with Berlekamp-Massey algorithm [Ber68]. The two limitations of RS codes are: the small block size and the high decoding times. The length of a RS code is limited by the field size, for example, it is $n \leq 255$ for $GF(256)$. The larger Galois field is, the longer the code can be, but at the same time the operations are getting slower and more complex. Therefore, the values of $n$ and $k$ have to be small if high transmission rates are desired.

The introduction of Turbo codes [BGT93] and LDPC codes [Gal63, MN97, MN95] have been one of the most important milestones in channel coding during the last years. Provided that the information block length is long enough, both Turbo and LDPC codes achieve performance close to the Shannon theoretical limit. However, in practical applications, both schemes have some complexity issues. Specifically, the encoding complexity is very low, but the decoding is more complex with Turbo codes compared to LDPC codes. The contrary occurs for standard LDPC codes. The encoding is more complex with standard LDPC codes, but the decoding is simpler.

LDPC codes were first introduced by Gallager in 1960 [Gal63], but they were impractical for implementation in that time. As the computational power has increased, they become attractive for research and practical implementations. MacKay and Neal rediscovered them in 1995 [MN95]. LDPC codes are defined by sparse parity-check matrices. Sparse bipartite graphs are used to represent LDPC codes where one set of nodes, the variable nodes, corresponds to elements of the codeword and the other set of nodes, the check nodes, corresponds to the set of parity-check constraints which define the code. An example of a regular LDPC code where all nodes of the same type have the same degree is shown in Figure 2.4. The principle of using irregular graphs where the degrees of the variable and the check nodes can vary widely was introduced in [LMS$^+$97], and it was further studied in [LMSS01b, RSU01]. The degrees of each set of nodes are chosen according to some distribution. The decoding complexity for LDPC codes increases linearly with the block length. LDPC codes are asymptotically good and their recovery performance decreases for small blocks.

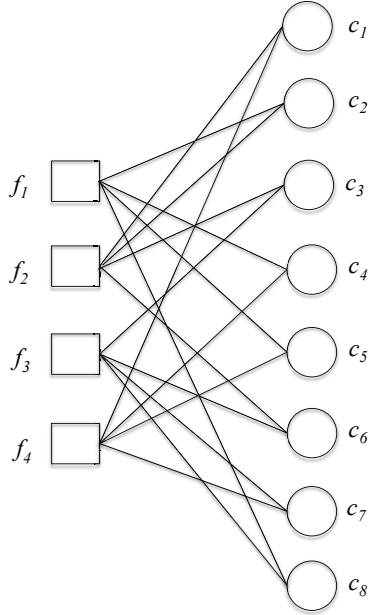

**Figure 2.4:** A $(2,4)$ regular LDPC code where $k = 8$ and $r = 4$. All 8 variable nodes have degree 2 and all 4 check nodes have degree 4.

Garcia-Frias and Zhong introduced Low-Density Generator Matrix (LDGM) codes that are constructed by using systematic sparse generator matrices [GFZ03]. They are a special class of LDPC codes with low encoding and decoding complexity. The amount of processing at the encoder is comparable to that of Turbo codes due to the sparseness of the generator matrix.

LDPC were significantly improved by Luby, Shokrollahi et al. that led to the invention of Tornado [BLMR98], LT [Lub02] and Raptor [Sho06] codes. Tornado codes are the precursor to fountain codes. Fountain or rateless codes are a class of erasure codes with the property that a potentially limitless sequence of encoding symbols can be generated from a given set of source symbols such that the original source symbols can ideally be recovered from any subset of the encoding symbols of size equal to or only slightly larger than the number of source symbols. These codes do not have a fixed code rate and are known as rateless codes. Both LT and Raptor codes belong to the class of rateless codes.

The fundamental goal of the research presented in the present thesis is to construct erasure codes that have the following desirable properties:

- High code rate;
- Same error-correcting capabilities as MDS codes, or very close to those of MDS codes; and
- Efficient encoding and decoding algorithms.

## 2.2 Network Coding

Network coding as a research area was initiated by the seminal paper by Ahlswede et al. [ACLY00]. They made the key observation that intermediate nodes are allowed to carry out algebraic operations on the incoming data instead of only forwarding the incoming data. Before defining their main result, some essential terminology that can be found in [MS12, Law01, Bol79, RSK10] is introduced.

A communication network is defined as a tuple $N = (V, E, S, T)$ that consists of:

- a finite directed acyclic multigraph $G = (V, E)$ where $V$ is the set of vertices and $E$ is the multiset of directed edge;
- a set $S \subset V$ of sources; and
- a set $T \subset V$ of sink nodes.

Vertices model communication nodes within the network, while directed edges model error-free communication channels between the nodes. An edge $(i, j)$ has unit capacity in the sense that it can be used to reliably deliver one symbol from $i$ to $j$. To allow for greater capacity from $i$ to $j$, multiple edges between $i$ and $j$ are permitted, i.e. $G$ is in general a multigraph. The capacity of an edge $(i, j) \in E$ is given by $R_{ij}$ and let $R = [R_{ij}, (i, j) \in E]$.

**Definition 2.10.** $F = [F_{ij}, (i,j) \in E]$ is a *flow* in $G$ from $s$ where $s \in S$ to $t_l$ where $t_l \in T$ if for all $(i, j) \in E$
$$0 \leq F_{ij} \leq R_{ij}$$
such that for all $i \in V$ except for $s$ and $t_l$
$$\sum_{i':(i',i) \in E} F_{i'i} = \sum_{j:(i,j) \in E} F_{ij},$$
i.e. the total flow into node $i$ is equal to the total flow out of node $i$.

$F_{ij}$ is referred to as the value of $F$ in the edge $(i, j)$. The value of $F$ is defined as
$$\sum_{j:(s,j) \in E} F_{sj} - \sum_{i:(i,s) \in E} F_{is}$$
which is equal to
$$\sum_{i:(i,t_l) \in E} F_{it_l} - \sum_{j:(t_l,j) \in E} F_{t_l j}.$$

**Definition 2.11.** $F$ is a *max-flow* from $s$ to $t_l$ in $G$ if $F$ is a flow from $s$ to $t_l$ whose value is greater than or equal to any other flow from $s$ to $t_l$.

**Definition 2.12.** A *cut* is a set of edges that partition the graph into two sets of vertices.

**Definition 2.13.** A *minimal cut* separating $s$ and $t$ is a cut of the smallest cardinality denoted as min-cut$(s, t)$.

A min-cut$(s, t)$ can be regarded as a bottleneck between $s$ and $t$ and it limits the information rate of the flow between $s$ and $t$.

**Theorem 2.14.** *[Law01]***Max-Flow Min-Cut Theorem:** *For every non-source node $t$, the minimum value of a cut between the source and a node $t$ is equal to max-flow$(s, t)$.*

It is well known from Menger's Theorem [Men] that the number of edge-disjoint paths from $s$ to $t$ is equal to max-flow. A collection of edge-disjoint paths can be found by the Ford-Fulkerson algorithm [FF]. Thus, in a single-sink network, the number of symbols transferred from $s$ to $t$ per time unit is equal to the min-cut of the network where each symbol is sent on a different edge-disjoint path.

Ahlswede et al. [ACLY00] showed that the multicast capacity for a single-source network, i.e. the maximum rate at which $s$ can transfer information to the sinks, cannot exceed the capacity of any cut separating $s$ from the sinks.

**Theorem 2.15.** *[MS12] The multicast rate $R(s, T)$ from $s$ to $T$ cannot exceed the transmission rate that can be achieved from $s$ to any element of $T$. The multicast rate $R(s, T)$ must satisfy:*

$$R(s, T) \leq \min_{t \in T} \text{min-cut}(s, t). \qquad (2.7)$$

The quantity $\min_{t \in T} \text{min-cut}(s, t)$ is referred to as the multicast capacity of the given network. The upper bound is achievable (with equality) via network coding [ACLY00].

Li et al. [LYC03] showed that Linear Network Coding (LNC) achieves the upper bound given that the finite field is sufficiently large.

**Theorem 2.16.** *Adapted version of [LYC03, Th.5.1] from [RSK10]: Let $q$ be a sufficient large power of 2. A symbol over $\mathbf{F}_q$ is treated as a unit of information. In a directed, delay-free, acyclic graph, with a single source $s$ and multiple sinks $t_1, \ldots, t_k$ and where the edges have integral capacity, if the capacity of the min-cut from the source to each of the $k$ sinks is at least $\nu$, then there exists a linear network solution that delivers $\nu$ units of information to each of the $k$ sinks simultaneously.*

Koetter and Médard extended further the work by Li et al. [LYC03] to arbitrary networks and robust networking. In [KM03], they presented an algebraic framework for investigating coding rate regions in networks using linear codes.

An efficient distributed randomized approach that asymptotically achieves the capacity for general multi-source multicast networks is presented in [HMK$^+$06]. Random Linear Network Coding (RLNC) is a simple, randomized coding method that maintains "a vector of coefficients for each of the source processes," which is "updated by each coding node". In other words, RLNC requires messages being communicated through the network to be accompanied by some degree of extra information (a vector of coefficients). The vector of coefficients is updated at each node that performs network coding.

Another definition of network coding is coding at a node in a packet network (where data is split into packets and network coding is applied to the content of packets). We use this definition in the sequel.

A well-known benefit of network coding is an increase of the throughput. This is achieved by using packet transmissions more efficiently, i.e. by communicating more information with fewer packet transmissions. The famous butterfly network in Figure 2.5 illustrates this. Assume that the source node $s$ wants to multicast two packets to the destinations $t_1$ and $t_2$. Assume that the capacity of each link is 1 packet per time unit and the delay of each link is the same. The maximum throughput from the source node $s$ to the destination nodes $t_1$ and $t_2$ is 2 packets per time unit, but the maximum throughput cannot be achieved simultaneously if only routing is allowed, since node $n_3$ can transmit either $b_1$ or $b_2$ but not both packets at the same time. Nevertheless, if node $n_3$ performs the exclusive-OR (XOR) operation on $b_1$ and $b_2$ and transmits the XOR-ed packet to node $n_4$, then both destinations achieve the maximum throughput simultaneously. Node $t_1$ decodes correctly $b_2$ after it receives $b_1$ from node $n_1$ and the XOR-ed packet from node $n_4$. It is similar for node $t_2$.

At the expense of encoding operations at the intermediate nodes and decoding operations at the sink nodes, RLNC improves the network throughput, the efficiency and the scalability, as well as the resilience to attacks and eavesdropping [CY02, BN05]. Inspired by these gains, researchers have applied network coding in many applications such as wireless networks [KRH$^+$08, KHH$^+$13], distributed storage systems [ADMK05, DGW$^+$10], video streaming [NNC10], satellite networks [VB09] and distributed file sharing [WWX10, FR12, GR06].

However, there are some issues with practical implementation of network coding. In order to explain them easily, the generation of a coded packet $y_k$ in RLNC is presented in Figure 2.6. The file is divided into $n$ packets of length $l$ and encoding is performed within a group of $m$ packets. This group is called a generation and $m$ is the generation size. Random coefficients are generated and each packet is multiplied with a coefficient. Then, all packets are XOR-ed together, i.e. bitwise XOR-ing of packets with equal length, and $y_k$ is generated. The newly generated packet is a linear combination of $m$ packets from the generation and each newly generated packet should be linearly independent from previously generated packets of the same generation. The average number of packets that have to be received before the original $m$ packets can be decoded is upper bounded by [LMS09]

$$\min\left\{m\frac{q}{q-1}, m+1+\frac{1-q^{-m+1}}{q-1}\right\}. \tag{2.8}$$

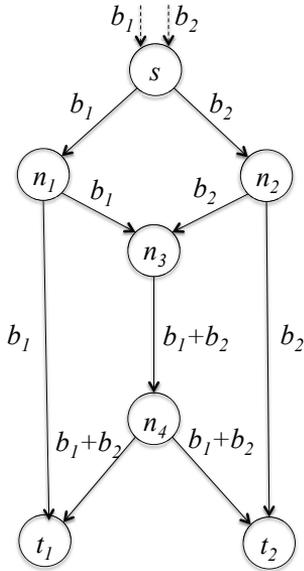

**Figure 2.5:** Butterfly network

The exact probability of successful decoding is derived in [TCBOF11]. The probability of generating a linearly independent packet increases with the number of packets in the generation $m$ and the size of the Galois field $q$. On the other hand, the length of the header overhead due to the coding coefficients becomes significant. This affects the throughput of a system and has a huge impact on the system load for some network scenarios. Thus, it is of a great importance to find a good tradeoff between the parameters.

Next, we calculate the encoding complexity of RLNC. The computational complexity of generating the coding coefficients depends on the complexity of generating a random number $r$ that is system specific and the generation size $m$. Consequently, the complexity of generating the encoding coefficients is $\mathcal{O}(mr)$ where $r$ is a constant. After generating the encoding coefficients, the packets are multiplied with them. This has a complexity of $\mathcal{O}(ml)$. Finally, the complexity of encoding one packet $y_k$ is $\mathcal{O}(m(l+r))$ [HPFL08], while encoding all packets within one generation has a complexity equal to $\mathcal{O}(m^2(l+r))$. The structure of a coded packet in RLNC is showed in Figure 2.7. The length of the vector of coding coefficients is $m \log_2 q$ bits.

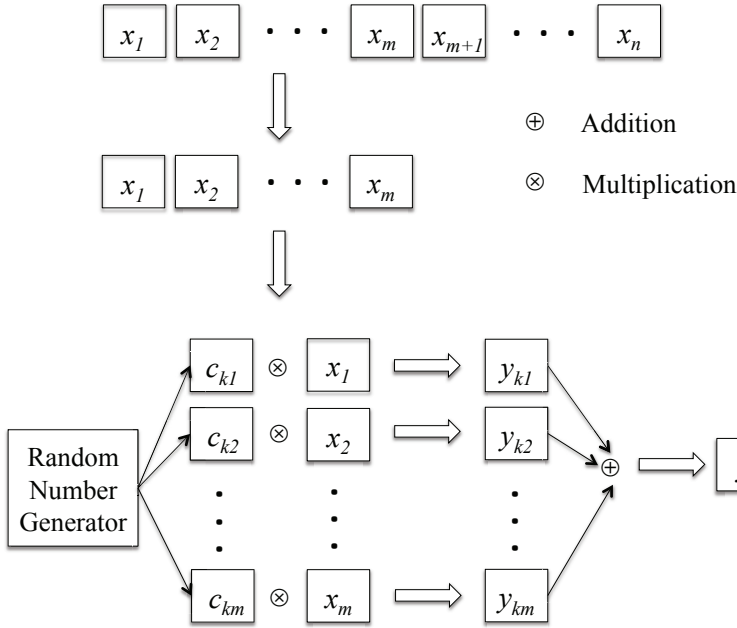

**Figure 2.6:** Generating a coded packet $y_k$ in RLNC. The file is split into $n$ packets and encoding is performed within a group of $m < n$ packets. Each packet is multiplied with a random coefficient $c_{k_i}$. All packets are XOR-ed together and $y_k$ is generated.

| Existing header | Vector of coding coefficients | Coded payload |
|---|---|---|

**Figure 2.7:** Structure of a coded packet in RLNC.

For instance, a typical packet length in sensor networks is 30 bytes. Consider a sensor network where 60 nodes send data. If RLNC in $GF(16)$ is performed and the generation size is 60, then 30 bytes per packet are used for recording the coding coefficients, i.e. the length of the header overhead is equal to the length of the useful data. Additionally, the header overhead has an impact on the required energy to transmit the coded packets. The energy used to transmit a single bit of data between devices in ad hoc sensor networks is equal to the energy for performing 800 instructions in the devices [MFHH02]. This implies that many applications may benefit by performing local computations rather than sending more bits. Thus, the reduction of the length of transmitted data, while keeping the same functionality of the employed algorithms is a challenging task.

Several header compression algorithms have been suggested in recent literature. Kötter and Kschischang proposed an approach that finds a linear subspace of the ambient vector space, and the coding is just done in that linear subspace [KK08]. This is a challenging task since every combination of source data should result in a distinct union subspace and finding a proper subspace can be a computational challenge.

The concept of sparse coding is well known, and it was first applied for header compression in network coding by Siavoshani et al. [SKFA09]. The number of combined packets in one coded packet is reduced from $n$ to $m$, where $m < n$, which results in a header length of $\mathcal{O}(m \log_2 n \log_2 q)$ bits. However, limiting the number of combined packets affects the invertibility of the matrix or it reduces the probability of a redundant packet being innovative [BKW97, HPFM11, FLS$^+$14, PFS05]. It was proved that $m$ should be at least of order $\mathcal{O}(\log n)$ so that the matrix has a full rank with high probability.

A header compression algorithm based on erasure decoding and list decoding was presented in [LR10]. The compressed header length under the erasure decoding scheme is $m + n/\log_2 q$ bits. The header length becomes arbitrarily close to $m + O(\log_2 n)/\log_2 q$ bits when the list decoding scheme is used. Both schemes are valid for moderate or large values of $m$.

In [CCW10], the header overhead is the seed for generating the coding coefficients with a known Pseudo-Random Number Generator (PRNG). This effectively reduces the header overhead to the size of the seed, but it does not support re-encoding which is the crucial constituent of RLNC [LWLZ10].

A similar solution where the generation of the coding coefficients is based on modified Vandermonde matrices which can be determined by one symbol is given in [TF12]. Two main drawbacks of this solution are: the network coding nodes can only perform addition operations and the generation size is upper bounded by $\log_2 q$ bits due to the cyclic property of the matrices.

Silva showed that precoding with Maximum Rank Distance (MRD) codes virtually eliminates the linear dependency even over a binary field [Sil12]. Coding in small finite fields significantly reduces the overhead in RLNC. This approach implies a moderate increase in the decoding complexity, but it potentially simplifies the operations at intermediate nodes that comes as an additional benefit besides minimizing the total overhead.

Recently, Fulcrum codes were proposed [LPHF14b, LPHF14a]. Fulcrum codes are concatenated codes where a seed for a PRNG is used to end-to-end communicate the coefficients of the outer code, while the inner code requires $1 + r/n$ bits per packet. Recoding can be performed exclusively over the inner code in $GF(2)$. Encoding and decoding is performed over the outer code in big finite fields.

Although there has been a vast amount of research results for network coding since its emergence, still there have not been many practical applications. The concept of network coding has been used to derive the bounds of the repair bandwidth in distributed storage systems. This is discussed in the next subsection.

The research in the present thesis addresses one of the main limitations for practical implementation of network coding:

– Reducing the length of the vector of coding coefficients.

## 2.3   Code Constructions for Distributed Storage Systems

A *distributed storage system* is a network of storage disks or nodes where data pertaining to a single file is distributed across the storage nodes. It is a practical choice for storing large amounts of data. The nodes are relatively inexpensive storage devices that may fail, be down during maintenance, or otherwise unavailable due to serving other demands, etc. A distributed storage system has to guarantee a reliable storage of the data over long periods of time even though the nodes might be individually unreliable.

**Definition 2.17.** *Reliability* is the probability that a system provides an uninterrupted service during a certain time interval $[0, t]$, i.e.

$$R(t) = P(T_{FF} > t), \qquad (2.9)$$

where $T_{FF}$ is Time to First Failure.

**Definition 2.18.** Mean Time to First Failure (MTFF) or Mean Time To Data Loss (MTTDL) is a measure of the reliability of a system defined as

$$MTFF = \int_0^\infty R(t)dt. \qquad (2.10)$$

It equals the time it takes a given storage system to exhibit enough failures such that at least one block of data cannot be retrieved or reconstructed.

In order to build a highly reliable system, redundancy has to be introduced. The redundancy can either be a simple copy of the data or a linear combination of the original data. *Replication* is a method of making copies from the original data. The data is available until one copy still exists. In case of storing one extra replica, the storage overhead is 100%, while it is 200% for 2 replicas and so forth. For instance, Google File System [GGL03] and Hadoop File System [SKRC10] store three copies of the data by default. When storing petabytes of data, replication is cost-inefficient. The main advantages of replication are: simple design and verification, low I/O and latency. However, its major disadvantage is the high storage overhead (200% for the industry standard) that translates into a high hardware cost (disk drives and associated equipment), as well as a high operational cost such as building space, power, cooling, maintenance, etc.

Weatherspoon and Kubiatowicz showed in [WK02] that erasure resilient systems use an order of magnitude less bandwidth and storage to provide a similar level of reliability as replicated systems. Hence many distributed storage systems are now turning to Reed-Solomon codes. Two reasons why often RS codes are employed in large-scale distributed systems are their storage optimality (since they are MDS codes) and generic applicability (construction of RS codes for arbitrary $n$ and $k$). Redundant Arrays of Inexpensive Disks (RAID) is a well known technology for data protection in high performance computing storage systems [PGK88]. RAID based systems use RS codes to recover the data when multiple disks fail simultaneously. In order to avoid the large finite field operations, many code constructions such as EvenOdd [BBBM95], Row-Diagonal Parity [CEG$^+$04] and flat XOR [GLW10] use only exclusive-OR operations to recover the data. Although RS codes improve the storage efficiency, the amount of accessed and transferred data to repair a failed node is large. We graphically present the repair process of a single systematic node with RS code in Figure 2.10 (a). The parameters in this example are $k = 10$, $r = 4$ and $M = 100$MB. When RS codes are used, each node is recovered by transferring data of size $M/k$ from any $k$ nodes. In order to reconstruct 10MB of data stored in a single node, $10 \times 10$MB=100MB are read from 10 nodes and transferred across the network to the node performing the decoding computations. Accordingly, RS codes perform poorly in terms of the repair bandwidth since the amount of the repair bandwidth is $k$ times the size of the data to be reconstructed. Additionally, this has a negative effect on the read performance of the system in a degraded mode (there is a read request for a data unit that is missing or unavailable) and the recovery time.

An improvement is to seek for codes that perform better than RS codes. Three major repair cost metrics for new erasure coding solutions have been identified in the recent literature: i) the amount of transferred data during a repair process (*repair bandwidth*), ii) the number of access operations during each repair (*disk-I/O*), and iii) the number of nodes that participate in the repair process (*repair locality*). Two types of repairs have been suggested[DRWS11]: *exact* (the recovered data has exactly the same content as the lost data) and *functional* (the newly generated data can be different than the lost one, but it maintains the MDS-property). The research presented in the present thesis focuses only on exact repair, which is preferred from a practical point of view.

### 2.3.1 Regenerating Codes

Classical MDS codes are optimal in terms of the storage-reliability tradeoff, but they still do not give a good answer to the key question: *How to encode the data in a distributed way while transferring as little data as possible across the network during a repair process?*

Under a $(n, k)$ MDS code, a file of size $M$ symbols is divided into $k$ fragments, each of size $\frac{M}{k}$ symbols, encoded and stored in $n$ nodes. The original file can be recovered from any set of $k$ fragments. Hence any data collector would preferably connect to these $k$ nodes. Dimakis et al. [DGW$^+$10] showed that the repair problem under functional repair can be mapped to a multicasting problem in an information flow graph [DGW$^+$10]. Building on known results from network coding, they proved that a node repair is possible if and only if the underlying information flow graph has sufficiently large min-cuts. Figure 2.8 shows an information graph that corresponds to a $(5, 3)$ code where the node $S$ corresponds to the source file. Assume that a $(5, 3)$ MDS code is used to encode the file $M = 3$MB. The code generates 5 fragments each of size $\alpha = 1$MB (stored in the nodes $x_1, \ldots, x_5$) with the property that any 3 can be used to reconstruct the original data. The storage node $x_i$ is represented by a storage input node $x_{in}^i$ and a storage output node $x_{out}^i$. The capacity of the edge between these two nodes is equal to the amount of stored data in node $x_i$. Data collectors connect to subsets of active nodes through edges with infinite capacity. When node $x_5$ fails, a newcomer $x_6$ reconstructs the lost data from $x_1, \ldots, x_4$. The question is: What is the minimum amount of information that has to be communicated? The new storage node $x_6$ downloads $\beta$MB from $d = 4$ active nodes. The min-cut separating the source and the data collector must be larger than $M = 3$MB for a reconstruction to be possible. In this example, the min-cut value is given by $1 + 3\beta$, implying that $\beta \geq 0.33$MB is sufficient and necessary.

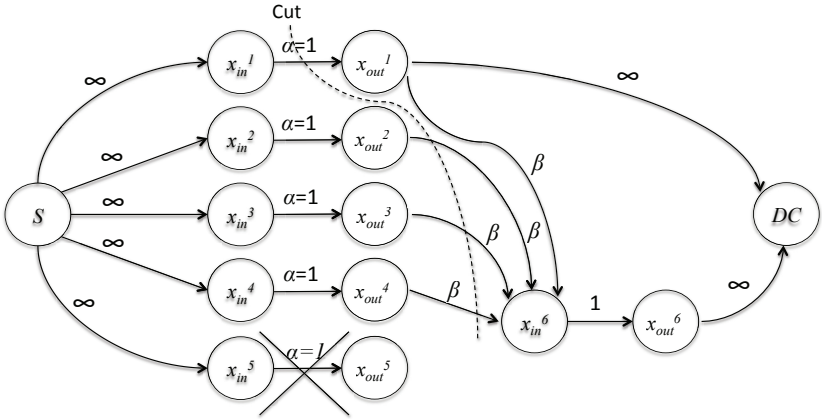

**Figure 2.8:** Illustration of an information flow graph corresponding to a $(5, 3)$ MDS code. When node $x_5$ is unavailable, a new node $x_6$ reconstructs the data by connecting to $d = 4$ available nodes and downloading $\beta$MB from each node.

A $(n, k, d)$ regenerating code replaces the data from a failed node by downloading $\beta$ symbols from $d$ non-failed nodes where $\beta \leq \alpha$. Thus, the repair bandwidth is $\gamma = d\beta$ where $\alpha \leq \gamma \ll M$. The main result of [DGW$^+$10] is the following Theorem.

**Theorem 2.19.** *[DGW$^+$10, Th.1] For any $\alpha \geq \alpha^*(n, k, d, \gamma)$, the points $(n, k, d, \alpha, \gamma)$ are feasible, and linear network codes suffice to achieve them. It is information theo-* retically impossible to achieve points with $\alpha < \alpha^*(n,k,d,\gamma)$. The threshold function $\alpha^*(n,k,d,\gamma)$ is the following:

$$\alpha^*(n,k,d,\gamma) = \begin{cases} \frac{M}{k}, \gamma \in [f(0), +\infty) \\ \frac{M - g(i)\gamma}{k-i}, \gamma \in [f(i), f(i-1)) \end{cases} \quad (2.11)$$

where

$$f(i) \triangleq \frac{2Md}{(2k-i-1)i + 2k(d-k+1)} \quad (2.12)$$

$$g(i) \triangleq \frac{(2d - 2k + i + 1)i}{2d} \quad (2.13)$$

where $d \leq n-1$. For $d, n, k$ given, the minimum repair bandwidth $\gamma$ is

$$\gamma_{min} = f(k-1) = \frac{2Md}{2kd - k^2 + k}. \quad (2.14)$$

The repair bandwidth $\gamma$ decreases as the number of $d$ nodes increases. An optimal tradeoff curve between the storage and the repair bandwidth for a $(15, 10, 14)$ code is shown in Figure 2.9. Regenerating codes achieve each point on this optimal tradeoff curve. The two extremal points on the optimal tradeoff correspond to the points at which the storage and the repair bandwidth are minimized, respectively. Following Theorem 2.19, the both points are attained when $d = n - 1$, i.e. all $n - 1$ non-failed nodes called helpers are contacted during a node repair. The codes that attain these points are known as *Minimum-Storage Regenerating (MSR)* and *Minimum-Bandwidth Regenerating (MBR)* codes, respectively. MSR codes achieve the pair

$$\left(\alpha_{MSR}, \gamma_{MSR}^{min}\right) = \left(\frac{M}{k}, \frac{M}{k}\frac{n-1}{n-k}\right), \quad (2.15)$$

while MBR codes

$$\left(\alpha_{MBR}^{min}, \gamma_{MBR}^{min}\right) = \left(\frac{M}{k}\frac{2n-2}{2n-k-1}, \frac{M}{k}\frac{2n-2.}{2n-k-1}\right). \quad (2.16)$$

MBR codes require an expansion factor of $\frac{2n-2}{2n-k-1}$ in the amount of stored data, thus, they are no longer storage-reliability optimal.

The work in [CJM10] and [SR10] showed that the lower bound of the repair bandwidth for functional repair with MSR codes given in (2.15) is also achieved for exact repair with MSR codes.

The present thesis focuses only on storage-reliability optimal codes with exact repair since they have the biggest practical potential.

We illustrate the benefit of using an MSR code in Figure 2.10 (b). We use the same parameters as for the RS code, i.e. $k = 10$, $r = 4$ and $M = 100$MB. Under the MSR code, the data from an unavailable node is reconstructed by downloading only $\frac{100}{40}$MB=2.5MB from each of the 13 non-failed nodes, i.e. $\frac{1}{4}$ of the stored data from

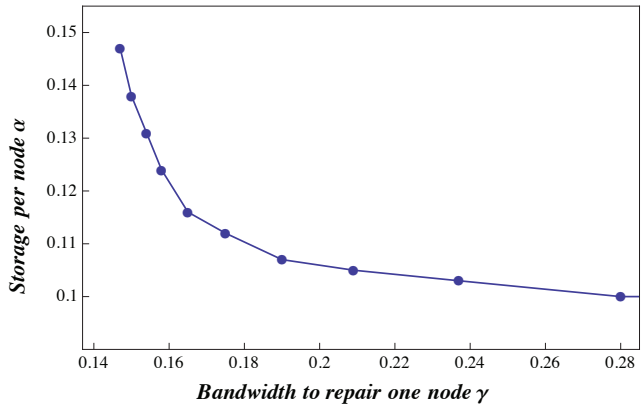

**Figure 2.9:** An optimal tradeoff curve between the storage $\alpha$ and the repair bandwidth $\gamma$ for a $(15, 10, 14)$ code and $M = 1$ [DGW$^+$10]. Traditional erasure coding (RS codes) corresponds to the points $\alpha = 0.1$ and $\gamma = 1$.

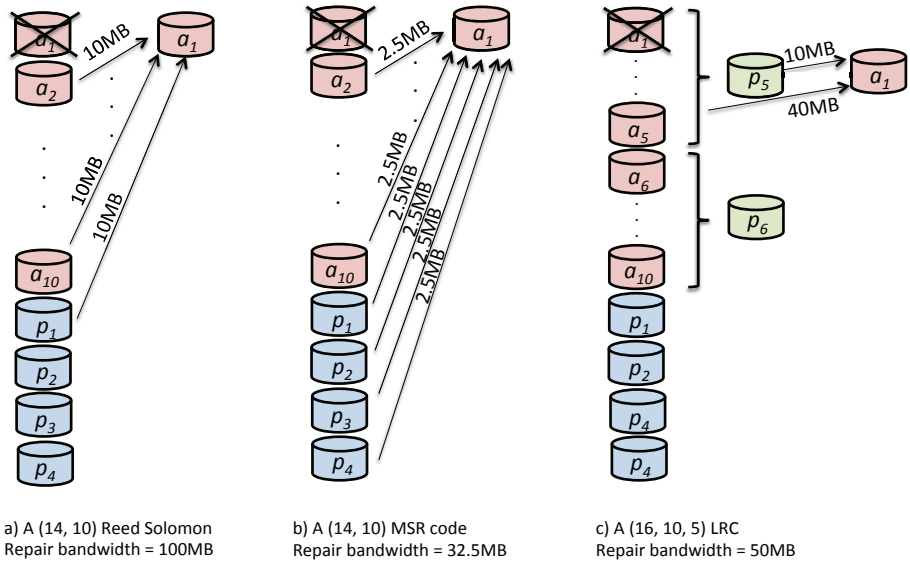

a) A (14, 10) Reed Solomon  
Repair bandwidth = 100MB

b) A (14, 10) MSR code  
Repair bandwidth = 32.5MB

c) A (16, 10, 5) LRC  
Repair bandwidth = 50MB

**Figure 2.10:** Amount of transferred data for reconstruction of the systematic node $a_1$ for a (14,10) RS code, a (14, 10) MSR code and a (16, 10, 5) LRC. The systematic nodes are represented in red and the parity nodes in blue, while the local parity nodes for the LRC are in green.

each node. The total repair bandwidth is only 32.5MB compared to 100MB under RS.

MSR codes possess all properties of an MDS code, giving an additional advantage of efficient repair consuming minimum possible bandwidth. This is made possible by using a sub-packetization level, i.e. the data at each node is further divided into blocks. The sub-packetization level $\alpha$ represents the minimum dimension over which all operations are performed. Namely, when the sub-packetization is 1, then each node is recovered by transferring data of size $M/k$ symbols from any $k$ nodes, i.e. the case when RS codes are used. Hence, $\alpha > 1$ is required to achieve the lower bound of the repair bandwidth. We use a sub-packetization level equal to 64 in Figure 2.10 (b) in order to achieve the minimum repair bandwidth with the $(14, 10)$ MSR code.

The construction of exact-repair codes is a well-studied problem in the literature. Here we only list the most relevant results for our work about MDS codes for storage systems. Exact-MSR codes that are obtained by using the technique of interference alignment, a technique used to efficiently handle multiple interfering channels in wireless communications [CJ08], were presented in [WD09, SR10].

Papailiopoulos et al. in [PDC13] and Cadambe et al. in [CHL11] resolved partly the open problem about designing high-rate MDS codes that achieve the optimal repair bandwidth. The code construction in [PDC13] used Hadamard matrices to construct a two-parity MDS code with optimal repair properties for any single node failure, including the parities. The construction is similar to zigzag codes in [TWB13], but the former one uses bigger finite fields. The zigzag codes provide an optimal recovery of any systematic node and an optimal update for a sub-packetization level of $r^k$. The work was further extended to provide an optimal recovery of both systematic and parity nodes [WTB11].

Furthermore, Tamo et al. [TWB14] showed that the sub-packetization level of an access-optimal MDS code for repairing a failed systematic node is $r^{\frac{k}{r}}$. In [CHLM11], Cadambe et al. proposed codes that repair optimally any systematic node for this sub-packetization level. An alternate construction of access-optimal MSR codes motivated by zigzag codes was presented in [ASVK15]. An essential condition for designing alternate codes with $r$ parity nodes is $m = \frac{k}{r}$ to be an integer $m \geq 1$ where $k$ is set to $rm$ and $\alpha$ to $r^m$.

Wang et al. constructed codes that optimally repair any systematic or parity node and require a sub-packetization level of $r^{k+1}$ [WTB11]. High-rate MSR codes with polynomial sub-packetization level were proposed in [SAK15]. However, the work presented in the present thesis focuses only on optimal repair of any systematic node.

Although the aforementioned MSR constructions achieve the lower bound of the repair bandwidth for a single failure, they have not been practically implemented in real-world distributed storage systems. Two main reasons for practical abandonment of existing MSR codes are: either MSR codes require encoding/decoding operations over an exponentially growing finite field or the sub-packetization level $\alpha$ increases exponentially. There are at least two ways to solve the problem by using small sub-packetization levels and optimizing in terms of I/O operations.

Rashmi et al. reported 35% reduction in the repair bandwidth for any systematic node when the sub-packetization level is 2 compared to a $(14, 10)$ RS code [RSG$^+$14]. They used the piggyback framework [RSR13] to construct Hitchhiker erasure codes. The basic idea of the piggyback framework is to take multiple instances of an existing code and add carefully designed functions of the data from one instance to another. Other code constructions with small sub-packetization levels are Rotated-RS [KBP$^+$12], EVENODD [BBBM95] and RDP codes [CEG$^+$04]. Rotated-RS codes exist only for $r \in \{2,3\}$ and $k \leq 36$, while EVENODD and RDP codes exist for $r = 2$.

I/O is becoming the primary bottleneck in the performance of cloud storage systems and applications that serve a large number of user requests or perform data intensive computations such as analytics. There are two main types of I/Os: sequential and random operations. *Sequential operations* access locations on the storage device in a contiguous manner, while *random operations* access locations on the storage device in a non-contiguous manner. A recent algorithm to transform Product-Matrix-MSR codes [RSK11] into I/O optimal codes (termed PM-RBT codes) while retaining their storage and network optimality was presented in [RNW$^+$15]. PM-RBT exist only for $r \geq k - 1$, i.e. for the low-rate regime.

All MDS erasure codes discussed in the previous paragraphs focus on an efficient repair from a single failure, since single failures present 98.08% of the total failures [RSG$^+$14]. On the other hand, the authors in [FLP$^+$10] state that the failures are often correlated. Next we review codes that outperform the aforementioned codes when multiple failures happen.

A cooperative recovery mechanism in the minimum-storage regime for repairing from multiple failures was proposed in [HXW$^+$10, WXHO10]. Minimum Storage Collaborative Regenerating (MSCR) codes minimize the repair bandwidth while still keeping the MDS property by allowing new nodes to download data from the non-failed nodes and the new nodes to exchange data among them. The repair bandwidth for MSCR codes under functional repair was derived independently in [HXW$^+$10, KSS11]. The existence of a random linear strong-MDS code under the assumption that the operations are in a sufficiently large finite field was showed in [HXW$^+$10]. The codes attain the MSR point but the decoding complexity is quite expensive. Adaptive regenerating codes where the numbers of failed and surviving nodes change over time were proposed in [KSS11]. The authors in [LL14] showed that it is possible to construct exact MSCR codes for optimal repair of two failures directly from existing exact MSR codes. MSCR codes that cooperatively repair any number of systematic nodes and parity nodes or a combination of one systematic and one parity node were presented in [CS13]. However, the code rate of these codes is low ($n = 2k$). A study about the practical aspects of codes with the same code rate ($n = 2k$) in a system called CORE that supports multiple node failures can be found in [LLL15]. There is no explicit construction of high-rate MDS codes for exact repair of multiple failures at the time of writing of the present thesis.

Practical scenarios [SAP+13, PJBM+16, RSG+14] showed that erasure codes have to provide a satisfactory tradeoff between the metrics such as storage overhead, reliability, repair bandwidth, locality and I/Os. The present thesis tackles this by working on:

– MDS code constructions with arbitrary sub-packetization levels for both low-rate and high-rate regimes; and

– Locally Repairable Codes.

### 2.3.2 Locally Repairable Codes

Locally Repairable Codes (LRCs) were independently introduced in [GHSY12, OD11, PLD+12].

Let $C$ be a $(n, k, d)_q$ linear code. Assume that the encoding of $\mathbf{x} \in \mathbf{F}_q^k$ is by the vector

$$C(\mathbf{x}) = (\mathbf{c}_1 \cdot \mathbf{x}, \ldots, \mathbf{c}_n \cdot \mathbf{x}) \in \mathbf{F}_q^n. \tag{2.17}$$

Thus, the code $C$ is specified by the set of $n$ points $C = \{\mathbf{c}_1, \ldots, \mathbf{c}_n\} \in \mathbf{F}_q^k$. The set of points must have a full rank equal to $k$ for $C$ to have $k$ information symbols.

**Definition 2.20.** For $\mathbf{c}_i \in C$, we define $Loc(\mathbf{c}_i)$ to be the smallest integer $l$ for which there exists $L \subseteq C$ of cardinality $l$ such that

$$\mathbf{c}_i = \sum_{j \in L} \lambda_j \mathbf{c}_j. \tag{2.18}$$

We further define $Loc(C) = max_{i \in \{1, \ldots, n\}} Loc(\mathbf{c}_i)$.

**Definition 2.21.** We say that a code $C$ has information locality $l$ if there exists $I \subseteq C$ of full rank such that $Loc(\mathbf{c}) \leq l$ for all $\mathbf{c} \in I$.

Gopalan et al. derived the upper bound for the minimum distance of a $(n, k, d)_q$ code with locality $l$.

**Theorem 2.22.** *[GHSY12, Th.5] For any $(n, k, d)_q$ linear code with information locality $l$,*

$$d \leq n - k - \left\lceil \frac{k}{l} \right\rceil + 2. \tag{2.19}$$

Huang et al. showed the existence of Pyramid codes that achieve the distance given in (2.19) when the field size is big enough [HCL13].

Two practical LRCs have been implemented in Windows Azure Storage [HSX+12] and HDFS-Xorbas by Facebook [SAP+13]. Both implementations reduce the reconstruction cost by introducing $l$ local and $r$ global parity blocks. The local parity nodes are computed from a subset of the systematic nodes. The locality of a code has also an impact on the fault tolerance and the update efficiency. LRCs tolerate $r+1$ *arbitrary node failures* and $l+r$ *theoretically decodable failures*. Consider the example with 10 data nodes in Figure 2.10 (c) where a (16, 10) LRC generates 6 (instead of 4) parity nodes. The first four parities (denoted as $p_1, p_2, p_3, p_4$) are global parities and are computed from all systematic nodes. While, for the two other parities, LRC divides the systematic nodes into two equal size groups and computes one local parity node for each group. The local parity $p_5$ is computed as an XOR combination from 5 systematic nodes in the first group $(a_1, \ldots, a_5)$, while the parity $p_6$ is computed from 5 systematic nodes in the second group $(a_6, \ldots, a_{10})$.

Let us consider the reconstruction of $a_1$. Instead of reading $p_1$ (or another global parity node) and the remaining 9 systematic nodes, it is more efficient to read $p_5$ and 4 systematic nodes $(a_2, \ldots, a_5)$ from the first group. It is easy to verify that the reconstruction of any systematic node requires accessing only 5 nodes, i.e. significantly less than the RS (10 nodes) and the MSR (13 nodes) code. The locality in this example is equal to 5 for the local parities and equal to 10 for the global parities. Any systematic node is recovered from $\frac{k}{l}$ nodes within its local group. Hence, the repair bandwidth for a single systematic node recovery with the (16, 10) LRC is 50MB. If we consider a repair of the parity nodes as well, then the average repair bandwidth for a single node failure is $(5 \times 12 + 10 \times 4) \times 10\text{MB}/16 = 62.5\text{MB}$, because the recovery of the global parities is performed in the same way as with RS codes.

The authors of [SAP+13] improved the recovery of the global parities by introducing an implied parity, but choosing the coefficients for the parities to satisfy the alignment condition is computationally demanding. Tamo et al. introduced a new family of optimal LRCs that are based on re-encoding RS encoded fragments [TPD13]. Although the code construction is simple, it requires a large finite field.

Motivated by practical situations in hot storage, where the data changes dynamically, the metric *update complexity* has been introduced in [ASV10]. Specifically, if the value of any of the systematic data changes, then the corresponding data has to be updated in the nodes that contain it in order to keep the data consistent.

**Definition 2.23.** The *update complexity* of a $(n, k, d)_q$ code $C$ is defined as the maximum number of symbols that must be updated when any single information symbol is changed.

Tolerating and recovering efficiently from multiple failures is an important requirement for big data storage systems. Shingled erasure codes (SHEC) are codes with local parities shingled with each other that provide efficient recovery from multiple failures [MNS14]. All parities have the same locality $l$ and support $\frac{rl}{k}$ systematic or parity node failures without data loss, but they are not efficient in terms of storage overhead and reliability.

There is still a need for new erasure resilient codes that reduce the number of nodes contacted during a repair, while still guaranteeing a low repair bandwidth even when multiple failures occur. Additionally, performing updates consumes bandwidth and energy, so it is of a great interest to construct codes that have small update complexity, i.e. small locality. Thus, one of the research topics in the present thesis is:

– A construction of LRCs that have low storage overhead, low average repair bandwidth for single and double failures, high reliability and improved update performance.

## 2.4   Optical Packet Switched Networks

*Dense Wavelength Division Multiplexing (DWDM)* has emerged as a core transmission technology for backbone networks. With DWDM, optical signals are multiplexed enabling simultaneously different wavelengths on the same fiber. Fiber networks can therefore carry multiple Terabits of data per second over thousands of kilometers [htt]. ADVA optical networking reported that the current DWDM systems support up to 192 wavelengths on a single fiber, with each wavelength transporting up to 100Gbit/s – 400Gbit/s and 1 Terabit/s.

However, DWDM opaque networks use expensive optical/electrical/optical (O/E/O) conversion for switching. Although the speed of electronic devices has been increased significantly, it is still not likely to catch up with the transmission speed available at the optical layer. Thus, there is a need to minimize or eliminate the electronic processing in order to fully exploit the potential bandwidth offered by DWDM. This calls for a move of the switching functionalities from the electronic domain to the optical domain, i.e. all-optical networking [OSHT01].

Furthermore, the traffic has become more data-dominant than voice-dominant. As the traffic nature has changed from continuous to bursty, there is a need for a switching technology that supports efficiently bursty traffic [QY99].

The switching technologies for DWDM are:

– *Wavelength Routed Optical Switching* establishes all-optical circuit switched connections (lightpaths) between edge nodes in the optical core network [RS02]. In Wavelength Routed Optical Networks (WRON), a lightpath is set-up before the data transmission and a dedicated wavelength on every link is reserved. The lightpath may be wavelength converted at the intermediate nodes. Thus, WRON does not require buffering, O/E/O conversion or processing at intermediate nodes. The major problem in WRON is the non-optimal utilization of link resources, because there is no resource sharing between lightpaths traversing the same link.

**Table 2.1:** A comparison of switching technologies for DWDM [VCR00].

| Switching technology | Granularity | Utilization | Complexity |
|---|---|---|---|
| Wavelength Switching (WRON) | Coarse | Poor | Low |
| Optical Packet Switching (OPS) | Fine | High | High (not mature) |
| Optical Burst Switching (OBS) | Moderate | Moderate | Moderate |

– An alternative to Wavelength Routed Optical Switching is *Optical Packet Switching (OPS)* and *Optical Burst Switching (OBS)* [GRG+98, YSH+98] that enable all-optical packet transport combined with statistical multiplexing for increased link utilization [HA00, CQY04, Tur99].

One of the main differences between OPS and OBS is the data unit that is processed and forwarded through the network. In OPS networks, packets are processed and forwarded, while in OBS networks, incoming packets are aggregated into bursts at an OBS ingress node based on the destination and/or the service class of the packets which are then transmitted in the network. Both OPS and OBS allow switching of data in the optical domain. However, switching decisions are made in the electronic domain [NBG08] as optical processing is still an immature technology.

Another key difference between OPS and OBS is the control information. The control information in OPS is in-band, while it is out-of-band in OBS networks. In particular, reservation is not possible in OPS, because the header follows the rest of the packet. In OBS, the burst transmission is initiated shortly after the burst was assembled and the control packet was sent out. The wavelength is allocated only for the duration of a data packet/burst and can be statistically shared by packets/bursts belonging to different connections. Switching decisions are taken based on the packet header or the burst control packet that undergoes O/E conversion and electronic processing at the switches, while the packet/burst payload is optically switched.

OPS has finer granularity compared to OBS, but requires fast switching, header reading and reinsertion that increase the complexity and the cost at the switching node [NBG08].

An overview of the characteristics of the aforementioned switching technologies is given in Table 2.1 [VCR00].

A crucial issue in OPS/OBS networks is the *packet loss at the network layer* due to contentions. A *contention* occurs at a switching node when two or more packets are destined on the same output port, on the same wavelength, at the same time. Contending packets are dropped which leads to increased PLR. For instance, the default behavior is to drop one of the two packets that contend for the same wavelength at the same time. In Figure 2.11 (a), packet *b* is dropped. It is impossible to buffer the data or retransmit the dropped packets. First, the amount of data is enormous, and, second, there is no Random-Access Memory (RAM) available in OPS/OBS networks. Thus, contending packets cannot be buffered and forwarded when the output port is free.

Several contention resolution mechanisms such as wavelength conversion, Fiber Delay Line (FDL) buffering and deflection routing [CWXQ03] have been proposed to combat packet loss in OPS/OBS networks. Wavelength conversion [DJMS98, EM00] converts the wavelength of one of the contending packets to an idle wavelength on the same output port (Figure 2.11 (b)). FDLs [HCA98] try to mimic electronic buffering by delaying one of the packets in time and scheduling it to the intended wavelength when it is free (Figure 2.11 (c)). A large number of FDLs are needed to implement large buffer capacity and they may add an additional delay. Both wavelength conversion and optical buffering require extra hardware which may increase the system cost. Deflection routing is a multiple-path routing technique that routes the contending packets to other nodes, i.e. output ports on the same wavelength. The performance of deflection routing largely depends on the network topology. A big advantage is that any of these three techniques can be combined. Another contention resolution scheme in OBS networks is burst segmentation. Rather than dropping the entire burst during contention, only the overlapping segments are dropped [VJS02].

FEC has been recently applied in OPS/OBS networks to alleviate PLR (Figure 2.11 (d)) [YKSC01]. FEC is not a contention resolution scheme, which means it can be combined with wavelength conversion, fiber delay line buffering and deflection routing. Redundant packets are added to a set of data packets at the OPS ingress node and transmitted along with the original data packets to an OPS egress node. Data packets dropped due to a contention can be reconstructed at the OPS egress node by using excess redundant packets, leading to a potential reduced PLR. The Network Layer Packet Redundancy Scheme (NLPRS), introduced in [Ove04], reduces several orders of magnitude the end-to-end PLR due to contentions in an asynchronous OPS ring network with and without a wavelength conversion. The authors in [GA02] evaluate several topology-routing algorithms for deflection routing coupled with erasure coding. The results show the amount of redundancy that is needed in unstructured networks of switches. The forward redundancy mechanism proposed in [VZ06] significantly reduces the packet loss compared to a retransmission-based backward loss recovery mechanism without the need for large ingress electronic buffers or big retransmission delays. Under this mechanism, only the overlapping segments of the contenting bursts are dropped. The dropped segments of a burst can be recovered using the redundant packets at the OBS egress node that were sent in the forward direction, from the ingress to the egress node.

Apostolopoulos used multiple state video coding and path diversity for transferring video over a lossy packet network [Apo01]. *Path diversity* is defined as sending different subsets of packets over disjoint paths, as opposed to the default scenarios where the packets proceed along a single path. Path diversity provides better performance because the probability that all of the multiple paths are simultaneously congested is much less than the probability that a single path is congested. Nguyen and Zakhor combined a rate allocation algorithm from multiple senders to a receiver with FEC in order to minimize the probability of lost packets due to congestions in a bursty loss environment [NZ02]. The work was further extended in [NZ03] where they presented a scalable, heuristic scheme for selecting a redundant path between an ingress node and an egress node. The disjoint paths from a sender to a receiver are created by using a collection of relay nodes.

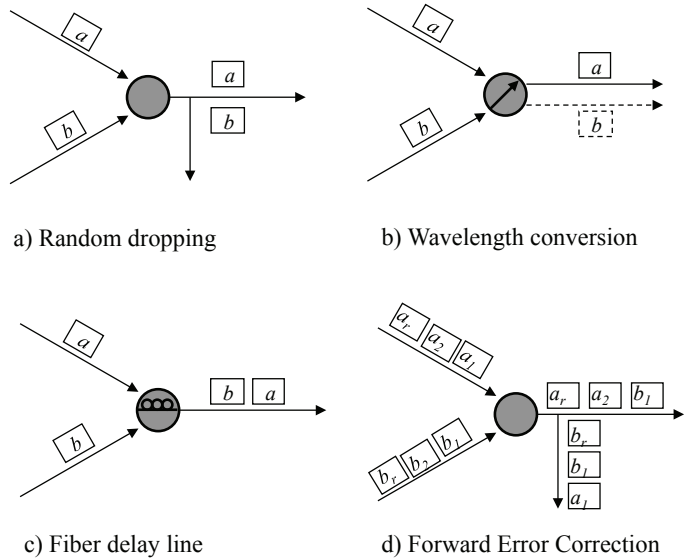

a) Random dropping  b) Wavelength conversion

c) Fiber delay line  d) Forward Error Correction

**Figure 2.11:** Contention resolution mechanisms at an OPS node where the packets *a* and *b* arrive on the same wavelength at the same time and contend for the same output wavelength. a) The packet *a* is transmitted, while *b* is dropped; b) Contention resolution with wavelength conversion where packet *b* is converted to an idle wavelength (on the same fiber); c) Contention resolution with FDL buffering where packet *b* is delayed using FDL buffering; d) Contention resolution with FEC where redundant packets are added.

Another coding technique for mitigating packet loss is network coding. A straightforward application of the network coding technique is not feasible in OPS/OBS due to the lack of store-and-forward capabilities. However, the idea of combining the packets instead of dropping them reduces the packet loss [BO11].

*Survivability* is essential in OPS/OBS networks with a throughput of order of terabits per second [ZS00]. The *survivability* of a network refers to a network's capability to provide continuous service in the presence of failures. The most common types of failures are node and link failure. Service providers have to ensure that their networks are fault tolerant. To meet these requirements, providers use common survivability mechanisms such as predesigned protection techniques [GL03, FV00, SRM02], node and component redundancy, prebuffering [KBOS10], and predetection schemes [BNH02]. The most used protection techniques include $1+1$ protection, in which the same data is transmitted on two link disjoint paths, and the receiver selects the packets from the path with better quality; $1:1$ protection, which is similar to $1+1$, except that traffic is not transmitted on the backup path until a failure occurs; $1:N$ protection, which is similar to $1:1$, except that one path is used to protect $N$ paths; and $M:N$, where $M$ protection paths are used to protect $N$ working paths. Figure 2.12 shows $1+1$ and $1:2$ path protection schemes.

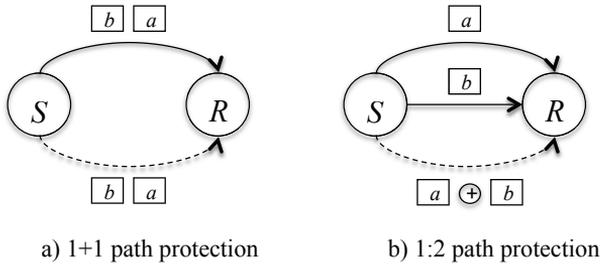

a) 1+1 path protection    b) 1:2 path protection

**Figure 2.12:** Different path protection schemes.

Some of the protection techniques can be combined with coding. Kamal first applied network coding to provide $1+N$ protection against single link failure [Kam06, Kam07b, Kam08]. The optimal $1+N$ protection scheme in [KAK08] requires exactly the same amount of protection resources as in $1:N$ and the time to recover from failures is comparable to that of $1+1$ protection. Kamal extended the approach to protect against multiple link failures [Kam07a, KRLL11]. Menendez and Gannet proposed photonic XOR devices for network coding [MG08]. Savings of up to 33% in links (transmitter, fiber or wavelength channel and receiver) are possible with network coding compared to conventional techniques. The effectivness of network coding to provide robustness against link failures of multicast traffic is presented in [MDXA10].

In OBS, data and control packets are sent out of band. Sending the Burst Header Packet (BHP) prior the transmission of data packets with specific offset time exposes the data payload to different security challenges. The authors in [SMS12] discussed the burst hijacking attack where a source node can maliciously create a copy of the original BCH and modify its value to setup a path to a malicious destination. The data payload is forwarded to the original destination as well as the malicious destination. However, the malicious destination does not send an acknowledgment for this hijacked burst, thus, it escapes from being caught. A solution based on Rivest-Shamir-Adleman (RSA) public-key encryption algorithm that addresses the Data Burst Redirection (DBR) attack in OBS networks is proposed in [CAKSAL+15]. However, the high capacities of OPS/OBS networks make data encryption in OPS/OBS not feasible as the current computational resources do not match the required encryption processing demands. The encryption mechanisms have to be with low computational complexity, suitable for high-speed implementation and the majority of the header content should not be encrypted since the processing of the headers has to be at ultra high speed [CV08].

One way of providing a certain level of security in a non-cryptographic way is to utilize non-systematic coding. *Secrecy* as defined in [MS12] provides protection from a passive adversary that is not able to reconstruct the whole packet/burst set by eavesdropping on a single path. This property has not been so far exploited in OPS/OBS networks. The authors in [OLV+12] showed how secrecy in storage systems is provided even when an eavesdropper knows or can guess some of the missing information.

There is a lack of research work that gives a unified view of the performance metrics in OPS. The scheme proposed in [Ove08] and [OBBT12] provides a $1+1$ path protection in addition to the packet loss alleviation. In particular, the work in [OBBT12] shows that significant cost savings are achieved by using erasure codes compared to other approaches that provide 1+1 path protection. There is a need for:

– An integrated view of QoS in OPS/OBS networks that deals with survivability, packet loss alleviation and secrecy at the same time.

# Chapter 3
# Contributions and Concluding Remarks

## 3.1 Research Questions

The state-of-the-art in the four main research areas covered in the present thesis was reviewed in the previous Chapter. The detailed literature study brings forth the following research questions.

First, it is of utmost importance to identify the desired code properties explained in Subsection 2.1 when constructing erasure codes, properties such as non-MDS or MDS, binary or non-binary codes with as few as possible operations in large finite fields, structure of the generator matrix, generic applicability, fast encoding/decoding etc. Accordingly, it is meaningful to ask:

**RQ1** How can we construct balanced erasure codes?

The next goal is to address some of the challenges of a practical implementation of network coding. Network coding can increase the data throughput by an order-of-magnitude and also improve the robustness of existing networks. However, one of the main challenges is the header overhead imposed by the coding coefficients as explained in Subsection 2.2. Accordingly, there is an enormous need for new header compression algorithms and this poses the next research question:

**RQ2** How can we reduce the header overhead in network coding?

Network coding concepts helped to derive the storage-repair bandwidth tradeoff for single node recovery with regenerating codes in distributed storage systems. While MSR codes possess all properties of MDS codes and offer an additional advantage of efficient repair consuming the minimum repair bandwidth (Subsection 2.3.1), locally repairable codes disregard the MDS property and provide a low locality (Subsection 2.3.2). Thus, a more general question about code constructions that are optimal for some of the cost metrics in distributed storage systems is raised:

**RQ3** How can we construct efficient codes for distributed storage systems?

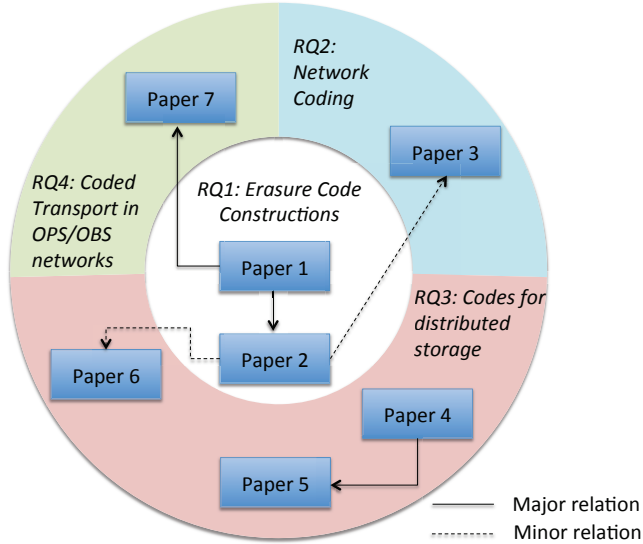

**Figure 3.1:** Relations between the papers included in the thesis. The papers are grouped based on the research questions.

The next question is related to provision of a unified view on QoS in OPS/OBS networks that is explained in Subsection 2.4. It is essential to know the interactions between the survivability, the packet loss rate and the secrecy when erasure codes are applied. Naturally, this poses the following question:

**RQ4** How can erasure coding be applied to increase the QoS in OPS/OBS networks?

## 3.2   Research Results

The author of the present thesis wrote and contributed to 10 publications and 7 patent applications during the four-year PhD period. Table 3.1 presents a complete list of included publications in the thesis, while Table 3.2 and Table 3.3 list the rest of the publications and patent applications that are not included in the thesis. The order in which the papers are given is not necessarily chronological, but rather related to the research questions so that it is easier and more natural to follow the exposition in the thesis.

Figure 3.1 shows the position of the included papers in the thesis with regards to the four research questions. Erasure code constructions are the core and they are used in different applications. The arrows depict the correlations between the included papers. Combinatorics and coding theory go hand in hand acting as powerful tools for design, analysis and implementation of efficient codes. We started with construction of erasure codes from combinatorial designs where we generate codes with balanced structure (Paper 1 and Paper 2). The balanced structure ensures that all packets have the same encoding complexity. A comparison between the decoding probability with the proposed balanced codes and RLNC is given in Paper 2. If the intermediate nodes in addition to the source nodes are allowed to encode the data, then we study a case of network coding. We identified the open problem of header compression that plays a major role in practical implementations of network coding.

**Table 3.1:** List of publications included in the thesis.

| Paper | Title · Author List · Conference/Journal |
|---|---|
| Paper 1 [KGO13] | **Balanced XOR-ed Coding** <br> K. Kralevska, D. Gligoroski, and H. Øverby <br> Lecture Notes in Computer Science, vol. 8115, pp. 161-172, 2013 |
| Paper 2 [GK14] | **Families of Optimal Binary Non-MDS Erasure Codes** <br> D. Gligoroski and K. Kralevska <br> IEEE Proceedings on International Symposium on Information Theory (ISIT), pp. 3150-3154, 2014 |
| Paper 3 [GKO15] | **Minimal Header Overhead for Random Linear Network Coding** <br> D. Gligoroski, K. Kralevska, and H. Øverby <br> IEEE International Conference on Communication Workshop (ICCW), pp. 680 - 685, 2015 |
| Paper 4 [KGO16b] | **General Sub-packetized Access-Optimal Regenerating Codes** <br> K. Kralevska, D. Gligoroski, and H. Øverby <br> IEEE Communications Letters, vol. 20, issue 7, pp. 1281 - 1284, 2016 |
| Paper 5 [KGJO17] | **HashTag Erasure Codes: From Theory to Practice** <br> K. Kralevska, D. Gligoroski, R. E. Jensen, and H. Øverby <br> Submitted to IEEE Transactions on Big Data |
| Paper 6 [KGO16a] | **Balanced Locally Repairable Codes** <br> K. Kralevska, D. Gligoroski, and H. Øverby <br> International Symposium on Turbo Codes and Iterative Information Processing, 2016 |
| Paper 7 [KOG15] | **Coded Packet Transport for Optical Packet/Burst Switched Networks** <br> K. Kralevska, H. Øverby, and D. Gligoroski <br> IEEE Proceedings on Global Communications Conference (GLOBECOM), pp. 1 - 6, 2015 |

**Table 3.2:** List of publications not included in the thesis.

| Paper | Title · Author List · Conference/Journal |
|---|---|
| Paper 8 [BCKO16] | **Combining Forward Error Correction and Network Coding in Bufferless Networks: a Case Study for Optical Packet Switching** <br> G. Biczók, Y. Chen, K. Kralevska, and H. Øverby <br> IEEE 17th International Conference on High Performance Switching and Routing (HPSR), 2016 |
| Paper 9 [AKJ16] | **Performance Analysis of LTE Networks with Random Linear Network Coding** <br> T. Degefa Assefa, K. Kralevska, and Y. Jiang <br> 39th International Convention on Information and Communication Technology, Electronics and Microelectronics, 2016 |
| Paper 10 [KOG14] | **Joint Balanced Source and Network Coding** <br> K. Kralevska, H. Øverby, and D. Gligoroski <br> 22nd Telecommunications Forum (TELFOR), Telecommunication Society, ISBN 978-1-4799-6190-0, pp. 589-592, 2014 |

**Table 3.3:** List of patent applications.

| Application No. | Title · Status |
|---|---|
| US14/902,251 [GK16] | **Network Coding over $GF(2)$** <br> Granted |
| PCT/EP2015/063337 [GK15, GK17a] | **Coding in Galois Fields with Reduced Complexity** <br> Pending |
| US 9,430,443 [JKGS16] | **Systematic Coding Technique** <br> Granted |
| GB1522869.5 [GK17b] | **Systematic Erasure Coding Technique** <br> Granted |
| GB1608441.0 | **Locally Repairable Erasure Codes** <br> Pending |
| GB1613575.8 | **Regenerating - Locally Repairable Codes** <br> Pending |
| GB1616704.1 | **Regenerating - Locally Repairable Codes** <br> Pending |

Employing some known facts from finite fields, we suggest an algorithm called Small Set of Allowed Coefficients (SSAC) in Paper 3. Network coding is an interesting problem that is well studied with the help of graph theory. By using some concepts of network coding, the lower bound of the repair bandwidth for a single node recovery in distributed storage networks is derived. A general construction of MDS codes for any sub-packetization level for repair of a single systematic node is suggested in Paper 4. The code construction from Paper 4 is further elaborated and optimized in terms of I/O in Paper 5. Later the idea of applying balanced codes as LRCs in distributed storage came. In Paper 6, we relaxed the condition of equal number of non-zero elements per column in the generator matrix. In both Paper 2 and Paper 6, we use hill-climbing search for codes with better performance. The path diversity and the secrecy features from Paper 1 inspired us to apply them in OPS/OBS networks in order to achieve survivability against link failures and non-cryptographic secrecy.

## 3.3 Research Answers

Here we summarize the answers to the research questions defined in Subsection 3.1.

**ANS1** Paper 1 and Paper 2 answer RQ1 by introducing a new way of constructing balanced binary codes from combinatorial designs (Latin squares and Latin rectangles).

**ANS2** Paper 3 answers RQ2 by proposing a novel algorithm called SSAC for practical network coding. SSAC generates the shortest header overhead by using sparse coding and properties of finite fields.

**ANS3** Paper 4, Paper 5 and Paper 6 answer RQ3. Paper 4 and Paper 5 present the first MDS codes for both low-rate and high-rate regimes that provide the lowest repair bandwidth for any sub-packetization level. The practicality of these codes is further investigated in Paper 5. Paper 6 solves partially the open problem of finding a general construction of LRCs with a low locality for any $n$ and $k$.

**ANS4** Paper 7 constitutes a first step for providing a unified view on the interactions between survivability, PLR and secrecy in OPS/OBS networks raised in RQ4.

## 3.4 Summary of the Results Contributing to the Thesis

This Section gives a summary of the papers included in Part II. The contributions of each paper are compared to the most relevant state-of-the-art results. The papers are presented in the order as they give answers to the research questions.

**Paper 1 [KGO13]: Balanced XOR-ed Coding**
K. Kralevska, D. Gligoroski, and H. Øverby
Lecture Notes in Computer Science, vol. 8115, pp. 161-172, 2013

Encoding and decoding over $GF(2)$ are up to two orders of magnitude less energy demanding and up to one order of magnitude faster than encoding/decoding operations in higher fields [VPFH10]. Sparse codes minimize the computation time of computing any coded packet, a property that is appealing to systems where computational load-balancing is critical. These are the main motivations to seek for sparse codes only with XOR operations.

The first theoretical work by Riis in [Rii04] shows that every solvable multicast network has a linear solution over $GF(2)$. Afterwards, XOR coding has been applied in wireless networks in [KRH+08] where random coding with no predefined code construction is used.

In Paper 1, we apply the knowledge from combinatorics for code constructions by introducing a new way of constructing balanced XOR-ed codes from combinatorial designs (Latin squares and Latin rectangles). In this context, balanced means all packets are encoded with equal complexity, i.e. the number of ones in each row and each column is equal. We show that the XOR-ed codes reach the max-flow for single source multicast acyclic networks with delay. Encoding of the original data is done by a nonsingular incidence matrix obtained from a Latin rectangle, while decoding is performed by the inverse matrix of the incidence matrix. Additionally, this paper shows that balanced coding offers plausible secrecy properties. In particular, if the incidence matrix and its inverse matrix switch the roles, then an eavesdropper has to eavesdrop at least max-flow links in order to decode one original packet.

**Paper 2 [GK14]: Families of Optimal Binary Non-MDS Erasure Codes**
D. Gligoroski and K. Kralevska
IEEE Proceedings on International Symposium on Information Theory (ISIT), pp. 3150-3154, 2014

The results presented in Paper 1 are further extended in Paper 2. We use the same logic (combinatorial designs) for constructing codes as in Paper 1. First, we define families of optimal binary non-MDS erasure codes. We also introduce the metric vector of exact decoding probability as a measure for how far away a specific $(n, k)$ code is from being an MDS code. The second contribution is a heuristic algorithm for finding those families by using hill climbing techniques over balanced XOR-ed codes. Due to the hill climbing search, those families of codes have always better decoding probability than the codes generated in a typical RLNC scenario. Finally, we show that for small values of $k$, the decoding probability of balanced XOR-ed codes in $GF(2)$ is very close to the decoding probability of random linear codes in $GF(4)$.

**Paper 3 [GKO15]: Minimal Header Overhead for Random Linear Network Coding**
D. Gligoroski, K. Kralevska, and H. Øverby
IEEE International Conference on Communication Workshop (ICCW), pp. 680 - 685, 2015

Paper 3 presents the only algorithm in the literature called SSAC where the header length does not depend from the size of the finite field. This is achieved by applying sparse coding, using an irreducible polynomial and at least two primitive elements from a finite field.

Although the concept of sparse coding is first used in [SKFA09], there the header length depends from the field size. Our work builds up on sparse coding, but we do not use parity-check matrices of error correcting codes. SSAC generates the header overhead by utilizing a small set $Q \subset GF(q)$ of coefficients that multiply the original data. Usually $Q$ consists of 2 primitive elements in $GF(q)$, thus, the header length is decreased from $n \log_2 q$ to $m(1 + \log_2 n)$ bits where $n$ is the generation size and $m$ is the sparsity parameter. We show that the header length in SSAC does not depend on the size of the finite field where the operations are performed, i.e. it just depends on the number of combined packets. Moreover, our work is the first one that investigates the efficiency of the header compression algorithm in every intermediate node in conjunction with the number of buffered packets in that node.

**Paper 4 [KGO16b]: General Sub-packetized Access-Optimal Regenerating Codes**
K. Kralevska, D. Gligoroski, and H. Øverby
IEEE Communications Letters, vol. 20, issue 7, pp. 1281 - 1284, 2016

In Paper 4, we propose an algorithm for explicit construction of MDS codes that access and transfer the lowest amount of data when repairing from a single node failure for any sub-packetization level. The amount of accessed and transferred data is the same. The number of helper nodes is $n-1$.

The algorithm presented in Paper 4 is so general that it also covers construction of access-optimal MSR codes. Compared to the construction presented in [ASVK15] where $\frac{k}{r}$ has to be an integer and the considered sub-packetization level is exclusively equal to $r^{\frac{k}{r}}$, the parameter $\frac{k}{r}$ in our work is not necessarily an integer. For instance, the code $(14, 10)$ that is deployed in the data-warehouse cluster of Facebook is out of the scope of applicability with the current proposals in [ASVK15, CHLM11, TWB14], because $\frac{k}{r} = 2.5$ is a non-integer. While the algorithm in Paper 4 constructs a $(14, 10, 13)$ code that reduces the repair bandwidth for any systematic node by $67.5\%$ when the sub-packetization level is $r^{\lceil \frac{k}{r} \rceil} = 64$ compared to a $(14, 10)$ RS code. The presented codes are simultaneously optimal in terms of storage, reliability and repair bandwidth. We also give an algorithm for exact repair of any systematic node that is linear and highly parallelized. This means a set of $\lceil \frac{\alpha}{r} \rceil$ symbols is independently repaired first and used along with the accessed data from other helper nodes to recover the remaining symbols. The results show how the repair bandwidth decreases as the sub-packetization level increases. The lower bound of the repair bandwidth is achieved for $\alpha = r^{\lceil \frac{k}{r} \rceil}$ (MSR codes).

**Paper 5 [KGJO17]: HashTag Erasure Codes: From Theory to Practice**
K. Kralevska, D. Gligoroski, R. E. Jensen, and H. Øverby
Submitted to IEEE Transactions on Big Data

Paper 5 is an extension of Paper 4 where we study both the theoretical and the practical aspects of the explicit construction introduced in Paper 4. Although we first introduced the codes without a specific name, in Paper 5 we call them *HashTag Erasure Codes (HTECs)* due to the resemblance between the hashtag sign # and the procedure of their construction. The three main contributions of Paper 5 are: an analysis of different concrete instances of HTECs, an elaboration of the correlation between the repair bandwidth and the I/Os with the sub-packetization level, and the repair bandwidth savings with HTECs even for repair of multiple failures.

We have implemented HTECs in C/C++ and performance analysis show up to 30% bandwidth savings compared to Piggyback 1 and Piggyback 2 codes [RSR13]. We also optimize HTECs in terms of the I/Os while still retaining their optimality in terms of the storage and the repair bandwidth. The authors in [RNW$^+$15] transform Product-Matrix-MSR (PM-MSR) into I/O optimal codes (which they call PM-RBT codes). Compared to HTECs that exist for any code parameters, PM-RBT codes exist only for $r \geq k - 1$. We identify the values of sub-packetization levels that give optimal overall system performance. We also show that the scheduling of the indexes in HTECs ensures a gradual increase in the number of random reads, hence no additional algorithms such as hop-and-couple [RSG$^+$14] are needed to make the reads sequential.

Additionally, HTECs are the first codes in the literature that offer bandwidth savings when recovering from multiple failures for any code parameters including the high-rate regime.

**Paper 6 [KGO16a]: Balanced Locally Repairable Codes**
K. Kralevska, D. Gligoroski, and H. Øverby
International Symposium on Turbo Codes and Iterative Information Processing, 2016

Paper 6 solves partially the open problem of finding a general construction of LRCs for any $n$ and $k$ [TPD13]. We suggest BLRCs that provide a good trade-off between the storage overhead, the repair bandwidth, MTTDL and the update complexity.

The main problem with many existing LRCs [HSX$^+$12, SAP$^+$13] is that although they reduce the size of the subset of contacted nodes, they suffer from the drawback that only a single subset of nodes enables the repair of a specific block. If a single node from that repair subset is not available, data cannot be repaired "locally" and this increases the repair cost. BLRCs address this problem and provide an efficient repair even when double failures occur. The strict requirement that the repair locality has to be a fixed small number $l$ is relaxed for BLRCs and we allow the repair locality to be either $l$ or $l + 1$. One of the main features of the proposed codes is that the parity blocks depend in a balanced manner from the systematic data blocks. This means that each systematic block is included in exactly $w$ parity blocks. Additionally, BLRCs are optimal even when double failures occur and this is not the case with other LRCs. We use four metrics such as storage overhead, average repair bandwidth, MTTDL and update complexity to compare our codes with existing LRCs. An extensive reliablity analysis for calculating the MTTDL is also presented.

**Paper 7 [KOG15]: Coded Packet Transport for Optical Packet/Burst Switched Networks**
K. Kralevska, H. Øverby, and D. Gligoroski
IEEE Proceedings on Global Communications Conference (GLOBECOM), pp. 1 - 6, 2015

Paper 7 provides a unified view on QoS in OPS/OBS networks. The work in Paper 7 focuses on the interactions between survivability, PLR and secrecy. The authors in [Ove04] and [VZ06] focus only on FEC codes to reduce packet loss in OPS networks. The work in [Ove08] and [OBBT12] extends these schemes to provide 1+1 path protection. Unlike our work, none of these references considers secrecy.

We present the Coded Packet Transport (CPT) scheme, a novel transport mechanism for OPS/OBS networks that exploits the benefits of source coding with erasure codes combined with path diversity. At an OPS/OBS egress node, reconstruction of lost packets due to contentions and link/node failures is enabled by the added redundancy. Sending different subsets of non-systematic coded packets over disjoint paths between the ingress and the egress node provides an end-to-end secrecy against passive adversaries. CPT provides a non-cryptographic secrecy in OPS networks. We combine an attack technique (a combination of partially known coded text and a brute force attack) with the modern recommended levels of security (a long term security level of 128 bits) to analyze the secrecy constraints in CPT. The presented analytical models show how the QoS aspects of CPT are affected by the number of disjoint paths, the packet overhead and the packet loss rate. The number of disjoint paths and the packet overhead should be chosen so that CPT is within the operational range (the secrecy and the survivability constraints are not violated).

## 3.5 Concluding Remarks

As the amount of generated and stored data is exponentially growing, cost-efficient and reliable systems have become increasingly important. One way to ensure efficient reliability are erasure codes with properties as close as possible to MDS codes. Erasure codes have become particularly attractive for fault protection in storage systems. In order to have a practical deployment of new erasure coding techniques in distributed storage, several issues have to be solved. New coding techniques have to provide high resilience to failures as well as low repair bandwidth and fast recovery.

Network coding offers throughput benefits, but at the cost of adding extra overhead. Transmitting a single bit in ad hoc sensor networks is more energy consuming than performing instructions in the devices. The large amount of traffic, has made all-optical network architectures crucial for high-speed transport. OPS is a promising candidate among all-optical network architectures proposed in recent literature. We elaborate the applicability of erasure coding in OPS/OBS networks in order to provide better QoS.

The present thesis has dealt with erasure code constructions, with a particular focus on general binary and non-binary code constructions and the advantages of employing them in different networks. The overall major scientific contributions in the present thesis include:

- A novel construction of binary codes suitable for implementation on devices with limited processing and energy capacity from combinatorial designs.

- A new algorithm for header compression in network coding where the header length is 2 to 7 times shorter than the length achieved by related compression techniques.

- Code constructions and implementation of new erasure codes for large scale distributed storage systems that provide savings in the storage and network resources.
  * A novel construction of MDS erasure codes that significantly reduce both the repair bandwidth and the random I/Os during a repair of missing or otherwise unavailable data with no additional storage overhead and flexibility in the choice of parameters.
  * A code construction optimized for repair locality and update complexity by relaxing the MDS requirement.

- A unified view of QoS in OPS/OBS networks by linking survivability, packet loss rate and secrecy using erasure coding and path diversity.

## 3.6  Future Works

Some proposals for future works are presented in this Section.

A natural follow-up of the current work related to large-scale distributed storage is an extension of the HTEC construction to enable construction of high-rate codes that are optimal for recovery of both the systematic and the parity nodes. Several high-rate MSR codes for efficient repair of both systematic and parity nodes exist in the literature [WTB11, SAK15, YB16a]. Still for these codes, either the sub-packetization level is too large or the constructions are not explicit. An open issue is how to extend the HTEC construction to support an efficient repair of the parity nodes as well.

In Paper 4 and Paper 5, we use the work in [ASVK15] to guarantee the existence of non-zero coefficients from $\mathbf{F}_q$ so that the code is MDS. However, the lower bound of the size of the finite field is relatively big. On the other hand, in our examples we actually work with very small finite fields ($\mathbf{F}_{16}$ and $\mathbf{F}_{32}$). Recent results in [YB16b] showed that a code is access-optimal for $\alpha = r^{\lceil \frac{k}{r} \rceil}$ over any finite field $\mathbf{F}$ as long as $|\mathbf{F}| \geq r \lceil \frac{k}{r} \rceil$. Determining the lower bound of the size of the finite field for HTECs remains an open problem.

Repairing the data with regenerating nodes requires contacting $n-1$ nodes. This creates a burden in the system and the data from all $n-1$ nodes should be always available. LRCs tackle this problem by contacting only $l$ nodes, but the improved performance comes at the expense of extra storage. An interesting research direction is how to combine the benefits from both regenerating and LRC codes.

Another research direction is the development of Error-Correcting Code (ECC) memory. The concept is similar to what we have been working on until now, but instead of recovering lost data distributed over different failure domains such as hard drives, nodes, racks and geographical locations, the concept is applied to memory chips. ECC memory offers error detection in addition to error correction. A memory error is an event that leads to the logical state of one or multiple bits being read differently from how they were last written. A memory error can lead to a machine crash or applications using corrupted data if the system does not support ECCs. Thus, ECCs memory are crucial components in each system. Since the construction of Hamming codes, only very few practical constructions have been designed and employed. Hamming codes are the most common used codes for protecting memory, although triple modular redundancy is used sometimes. Hamming codes are Single-Error-Correcting and Double-Error-Detecting (SECDED). BCH codes are another alternative for practical implementations of ECCs. They have better correcting capabilities than Hamming codes, but imply a higher latency. Thus, another research direction is a construction of low latency ECCs that offer multi-bit detection and correction in one cycle.

# References

[ACLY00]   R. Ahlswede, N. Cai, S. Y. R. Li, and R. W. Yeung. Network information flow. *IEEE Transactions on Information Theory*, 46(4):1204–1216, 2000.

[ADMK05]   S. Acedański, S. Deb, M. Médard, and R. Kötter. How good is random linear coding based distributed networked storage. In *NetCod*, 2005.

[AKJ16]   T. Degefa Assefa, K. Kralevska, and Y. Jiang. Performance analysis of lte networks with random linear network coding. In *2016 39th International Convention on Information and Communication Technology, Electronics and Microelectronics (MIPRO)*, pages 601–606, May 2016.

[Apo01]   J. G. Apostolopoulos. Reliable video communication over lossy packet networks using multiple state encoding and path diversity. In *VCIP*, Proceedings of SPIE, pages 392–409, 2001.

[ASV10]   N.P. Anthapadmanabhan, E. Soljanin, and S. Vishwanath. Update-efficient codes for erasure correction. In *48th Annual Allerton Conference on Communication, Control, and Computing*, pages 376–382, Sept 2010.

[ASVK15]   G.K. Agarwal, B. Sasidharan, and P. Vijay Kumar. An alternate construction of an access-optimal regenerating code with optimal sub-packetization level. *National Conference on Communications (NCC)*, pages 1–6, Feb 2015.

[BBBM95]   M. Blaum, J. Brady, J. Bruck, and Jai Menon. Evenodd: an efficient scheme for tolerating double disk failures in raid architectures. *Computers, IEEE Transactions on*, 44(2):192–202, Feb 1995.

[BCKO16]   G. Biczok, Y. Chen, K. Kralevska, and H. Overby. Combining forward error correction and network coding in bufferless networks: A case study for optical packet switching. In *2016 IEEE 17th International Conference on High Performance Switching and Routing (HPSR)*, pages 61–68, June 2016.

[Ber68]   E. R. Berlekamp. *Algebraic Coding Theory*. MacGraw-Hill (New York NY), 1968.

[BGT93]　C. Berrou, A. Glavieux, and P. Thitimajshima. Near shannon limit error-correcting coding and decoding: Turbo-codes. In *ICC*, volume 2, pages 1064–1070, 1993.

[BKW97]　J. Blömer, R. Karp, and E. Welzl. The rank of sparse random matrices over finite fields. *Random Struct. Algorithms*, 10(4):407–419, July 1997.

[BLMR98]　J. W. Byers, M. Luby, M. Mitzenmacher, and A. Rege. A digital fountain approach to reliable distribution of bulk data. Technical report, May 1998.

[BN05]　K. Bhattad and K.R. Narayanan. Weakly secure network coding. *Proc. First Workshop on Network Coding, Theory, and Applications (NetCod)*, 2005.

[BNH02]　S. Bjornstad, M. Nord, and D.R. Hjelme. Transparent optical protection switching scheme based on detection of polarisation fluctuations. In *Optical Fiber Communication Conference and Exhibit*, pages 433–434, Mar 2002.

[BO11]　G. Biczok and H. Overby. Combating packet loss in ops networks: A case for network coding. In *NIK*, 2011.

[Bol79]　B. Bollobas. *Graph Theory: An Introductory Course*. Springer, New York/Berlin, 1979.

[BRC60]　R. C. Bose and Dwijendra K. Ray-Chaudhuri. On A class of error correcting binary group codes. *Information and Control*, 3(1):68–79, March 1960.

[CAKSAL+15]　Y. Coulibaly, A. Al-Kilany, M. S. A. Latiff, G. Rouskas, S. Mandala, and M.A. Razzaque. Secure burst control packet scheme for optical burst switching networks. In *IEEE International Broadband and Photonics Conference (IBP)*, pages 86–91, April 2015.

[CC81]　G. C. Clark, and J. Bibb Cain. *Error-correction coding for digital communications*. Plenum Press, 1981.

[CC84]　R. Comroe and D.J. Costello. Arq schemes for data transmission in mobile radio systems. *IEEE Journal on Selected Areas in Communications*, 2(4):472–481, July 1984.

[CCW10]　C.-C. Chao, C.-C. Chou, and H.-Y. Wei. Pseudo random network coding design for IEEE 802.16m enhanced multicast and broadcast service. In *VTC Spring*, pages 1–5. IEEE, 2010.

[CEG+04]　Peter Corbett, Bob English, Atul Goel, Tomislav Grcanac, Steven Kleiman, James Leong, and Sunitha Sankar. Row-diagonal parity for double disk failure correction. In *Proceedings of the USENIX FAST '04 Conference on File and Storage Technologies*, pages 1–14, San Francisco, CA, March 2004. Network Appliance, Inc., USENIX Association.

[CHL11]   V. R. Cadambe, C. Huang, and J. Li. Permutation code: Optimal exact-repair of a single failed node in MDS code based distributed storage systems. In *2011 IEEE International Symposium on Information Theory Proceedings, ISIT 2011, St. Petersburg, Russia, July 31 - August 5, 2011*, pages 1225–1229. IEEE, 2011.

[CHLM11]  V.R. Cadambe, C. Huang, J. Li, and S. Mehrotra. Polynomial length mds codes with optimal repair in distributed storage. *Asilomar Conference on Signals, Systems and Computers*, pages 1850–1854, Nov 2011.

[CJ08]    V.R. Cadambe and S.A. Jafar. Interference alignment and degrees of freedom of the k -user interference channel. *IEEE Transactions on Information Theory*, 54(8):3425–3441, Aug 2008.

[CJM10]   V. R. Cadambe, S. A. Jafar, and H. Maleki. Distributed data storage with minimum storage regenerating codes - exact and functional repair are asymptotically equally efficient. *CoRR*, abs/1004.4299, 2010.

[CQY04]   Y. Chen, C. Qiao, and X. Yu. Optical burst switching (obs): A new area in optical networking research. *IEEE Network Magazine*, 18:16–23, 2004.

[CS13]    J. Chen and K. W. Shum. Repairing multiple failures in the suh-ramchandran regenerating codes. In *ISIT*, pages 1441–1445, 2013.

[CV08]    Yuhua Chen and P.K. Verma. Secure optical burst switching: Framework and research directions. *IEEE Communications Magazine*, 46(8):40–45, 2008.

[CWXQ03]  Y. Chen, H. Wu, D. Xu, and C. Qiao. Performance analysis of optical burst switched node with deflection routing. In *ICC*, pages 1355–1359. IEEE, 2003.

[CY02]    N. Cai and R.W. Yeung. Secure network coding. In *IEEE International Symposium on Information Theory*, pages 323–328, 2002.

[dB96]    M. A. de Boer. Almost mds codes. *Des. Codes Cryptography*, 9(2):143–155, 1996.

[DGW$^+$10] A.G. Dimakis, P.B. Godfrey, Y. Wu, M.J. Wainwright, and K. Ramchandran. Network coding for distributed storage systems. *IEEE Transactions on Information Theory*, 56(9):4539–4551, Sept 2010.

[DJMS98]  S.L. Danielsen, C. Joergensen, B. Mikkelsen, and K.E. Stubkjaer. Optical packet switched network layer without optical buffers. *Photonics Technology Letters, IEEE*, 10(6):896–898, 1998.

[DL95]    S.M. Dodunekov and I.N. Landjev. On near-mds codes. *Journal of Geometry*, 54:30–43, 1995.

[DRWS11]   Alexandros G. Dimakis, Kannan Ramchandran, Yunnan Wu, and Changho Suh. A survey on network codes for distributed storage. *Proceedings of the IEEE*, 99(3):476–489, 2011.

[DSDY13]   S. H. Dau, W. Song, Z. Dong, and C. Yuen. Balanced sparsest generator matrices for mds codes. In *IEEE International Symposium on Information Theory Proceedings (ISIT)*, pages 1889–1893, July 2013.

[EM00]   J.M.H. Elmirghani and H.T. Mouftah. All-optical wavelength conversion: technologies and applications in dwdm networks. *IEEE Communications Magazine*, 38(3):86–92, Mar 2000.

[FF]   L. R. Ford and D. R. Fulkerson. Maximal flow through a network. *Canadian Journal of Mathematics*, 8:399–404.

[FLP$^+$10]   D. Ford, F. Labelle, F. I. Popovici, M. Stokely, V.-A. Truong, L. Barroso, C. Grimes, and S. Quinlan. Availability in globally distributed storage systems. In *9th USENIX Symposium on Operating Systems Design and Implementation, OSDI*, pages 61–74. USENIX Association, 2010.

[FLS$^+$14]   S. Feizi, D.E. Lucani, C.W. Sorensen, A. Makhdoumi, and M. Medard. Tunable sparse network coding for multicast networks. In *International Symposium on Network Coding (NetCod)*, pages 1–6, June 2014.

[FR12]   M. H. Firooz and S. Roy. Data dissemination in wireless networks with network coding. *IEEE Communications Letters, Volume:17 , Issue: 5, 2013*, December 08 2012.

[FV00]   A. Fumagalli and L. Valcarenghi. Ip restoration vs. wdm protection: is there an optimal choice? *IEEE Network*, 14(6):34–41, Nov 2000.

[GA02]   S. Ghosh and V. Anantharam. Bufferless all-optical networking with erasure codes. In *IEEE Information Theory Workshop*, pages 19–24, Oct 2002.

[Gal63]   R. G. Gallager. *Low-Density Parity-Check Codes*. The M.I.T. Press, Cambridge, MA, USA, 1963.

[GFZ03]   J. Garcia-Frias and W. Zhong. Approaching shannon performance by iterative decoding of linear codes with low-density generator matrix. *IEEE Communications Letters*, 7(6):266–268, June 2003.

[GGL03]   S. Ghemawat, H. Gobioff, and S.-T. Leung. The google file system. In *ACM Symposium on Operating Systems Principles*, SOSP '03, pages 29–43. ACM, 2003.

[GHSY12]   P. Gopalan, C. Huang, H. Simitci, and S. Yekhanin. On the locality of codeword symbols. *IEEE Transactions on Information Theory*, 58(11):6925–6934, 2012.

[GK14]     D. Gligoroski and K. Kralevska. Families of optimal binary non-mds erasure codes. In *2014 IEEE International Symposium on Information Theory*, pages 3150–3154, June 2014.

[GK15]     D. Gligoroski and K. Kralevska. Coding in galois fields with reduced complexity, December 2015. WO Patent App. PCT/EP2015/063,337.

[GK16]     D. Gligoroski and K. Kralevska. Network coding over gf(2), December 2016. US Patent App. 14/902,251.

[GK17a]    D. Gligoroski and K. Kralevska. Coding in galois fields with reduced complexity, May 18 2017. US Patent App. 15/319,428.

[GK17b]    D. Gligoroski and K. Kralevska. Systematic coding technique for erasure correction, June 2017. WO Patent App. PCT/EP2016/082656.

[GKO15]    D. Gligoroski, K. Kralevska, and H. Overby. Minimal header overhead for random linear network coding. In *2015 IEEE International Conference on Communication Workshop (ICCW)*, pages 680–685, June 2015.

[GL03]     D. W. Griffith and S. Lee. A 1+1 protection architecture for optical burst switched networks. *IEEE Journal on Selected Areas in Communications*, 21(9):1384–1398, 2003.

[GLW10]    K.M. Greenan, X. Li, and J.J. Wylie. Flat xor-based erasure codes in storage systems: Constructions, efficient recovery, and tradeoffs. In *IEEE Symposium on Mass Storage Systems and Technologies (MSST)*, pages 1–14, May 2010.

[GR06]     C. Gkantsidis and P. Rodriguez. Cooperative security for network coding file distribution. In *INFOCOM*. IEEE, 2006.

[GRG+98]   P. Gambini, M. Renaud, C. Guillemot, F. Callegati, I. Andonovic, B. Bostica, D. Chiaroni, G. Corazza, S. L. Danielsen, P. Gravey, P. B. Hansen, M. Henry, C. Janz, A. Kloch, R. Krähenbühl, C. Raffaelli, M. Schilling, A. Talneau, and L. Zucchelli. Transparent optical packet switching: network architecture and demonstrators in the KEOPS project. *IEEE Journal on Selected Areas in Communications*, 16(7):1245–1259, 1998.

[HA00]     D.K. Hunter and I. Andonovic. Approaches to optical internet packet switching. *IEEE Communications Magazine*, 38(9):116–122, Sep 2000.

[Haf05]    J. L. Hafner. Weaver codes: Highly fault tolerant erasure codes for storage systems. In *FAST*. USENIX, 2005.

[HCA98]    D. K. Hunter, M. C. Chia, and I. Andonovic. Buffering in optical packet switches. *Journal of Lightwave Technology*, 16(12):2081–2094, Dec 1998.

[HCL13]   C. Huang, M. Chen, and J. Li. Pyramid codes: Flexible schemes to trade space for access efficiency in reliable data storage systems. *TOS*, 9(1):3, 2013.

[HLH16]   W. Halbawi, Z. Liu, and B. Hassibi. Balanced reed-solomon codes. In *IEEE International Symposium on Information Theory (ISIT)*, pages 935–939, July 2016.

[HMK$^+$06]   T. Ho, M. Medard, R. Kotter, D. R. Karger, M. Effros, J. Shi, and B. Leong. A random linear network coding approach to multicast. *IEEE Transactions on Information Theory*, 52(10):4413–4430, 2006.

[Hoc59]   A. Hocquenghem. Codes correcteurs d'Erreurs. *Chiffres (Paris)*, 2:147–156, September 1959.

[HPFL08]   J. Heide, M. V. Pedersen, F. H. P. Fitzek, and T. Larsen. Cautious view on network coding - from theory to practice. *Journal of Communications and Networks*, 10(4):403–411, 2008.

[HPFM11]   J. Heide, M. V. Pedersen, F. H. P. Fitzek, and M. Médard. On code parameters and coding vector representation for practical RLNC. In *ICC*, pages 1–5, 2011.

[HSX$^+$12]   C. Huang, H. Simitci, Y. Xu, A. Ogus, B. Calder, P. Gopalan, J. Li, and S. Yekhanin. Erasure coding in windows azure storage. In *USENIX Annual Technical Conference*, pages 15–26, 2012.

[htt]   http://www.advaoptical.com/en/products/technology/dwdm.aspx. *ADVA Optical Networking*.

[HXW$^+$10]   Y. Hu, Y. Xu, X. Wang, C. Zhan, and P. Li. Cooperative recovery of distributed storage systems from multiple losses with network coding. *IEEE Journal on Selected Areas in Communications*, 28(2):268–276, February 2010.

[IDC12]   IDC. Idc's digital universe study, sponsored by emc. *White Paper*, December 2012.

[JKGS16]   R.E. Jensen, K. Kralevska, D. Gligoroski, and S.B. Stene. Systematic coding technique, August 2016. US Patent 9,430,443.

[KAK08]   A.E. Kamal and O. Al-Kofahi. Toward an optimal 1+n protection strategy. In *Allerton Conference on Communication, Control, and Computing*, pages 162–169, Sept 2008.

[Kam06]   A. E. Kamal. 1+N protection in mesh networks using network coding over p-cycles. In *GLOBECOM*, 2006.

[Kam07a]   A. E. Kamal. 1+N protection against multiple link failures in mesh networks. In *ICC*, pages 2224–2229, 2007.

[Kam07b]   A. E. Kamal. GMPLS-based hybrid 1+N link protection over p-cycles: Design and performance. In *GLOBECOM*, pages 2298–2303, 2007.

[Kam08]    A. E. Kamal. A generalized strategy for 1+N protection. In *ICC*, pages 5155–5159, 2008.

[KBOS10]   A. Kimsas, S. Bjornstad, H. Overby, and N. Stol. Improving performance in the opmigua hybrid network employing the network layer packet redundancy scheme. *IET Communications*, 4(3):334–342, 2010.

[KBP+12]   O. Khan, R. C. Burns, J. S. Plank, W. Pierce, and C. Huang. Rethinking erasure codes for cloud file systems: minimizing I/O for recovery and degraded reads. In *Proceedings of the 10th USENIX conference on File and Storage Technologies, FAST 2012, San Jose, CA, USA, February 14-17, 2012*, pages 20–40. USENIX Association, 2012.

[KGJO17]   K. Kralevska, D. Gligoroski, R. E. Jensen, and H. Overby. Hashtag erasure codes: From theory to practice. *IEEE Transactions on Big Data*, PP(99):1–1, 2017.

[KGO13]    Katina Kralevska, Danilo Gligoroski, and Harald Overby. Balanced xor-ed coding. In *Advances in Communication Networking: 19th EUNICE/IFIP WG 6.6 International Workshop*, pages 161–172. Springer Berlin Heidelberg, 2013.

[KGO16a]   K. Kralevska, D. Gligoroski, and H. Overby. Balanced locally repairable codes. In *2016 9th International Symposium on Turbo Codes and Iterative Information Processing (ISTC)*, pages 280–284, Sept 2016.

[KGO16b]   K. Kralevska, D. Gligoroski, and H. Overby. General sub-packetized access-optimal regenerating codes. *IEEE Communications Letters*, 20(7):1281–1284, July 2016.

[KHH+13]   J. Krigslund, J. Hansen, M. Hundeboll, D. E. Lucani, and F. H. P. Fitzek. CORE: COPE with MORE in wireless meshed networks. In *VTC Spring*, pages 1–6, 2013.

[KK08]     R. Kötter and F. R. Kschischang. Coding for errors and erasures in random network coding. *IEEE Transactions on Information Theory*, 54(8):3579–3591, 2008.

[KM03]     R. Koetter and M. Médard. An algebraic approach to network coding. *IEEE/ACM Trans. Netw*, 11(5):782–795, 2003.

[KOG14]    K. Kralevska, H. Overby, and D. Gligoroski. Joint balanced source and network coding. In *Telecommunications Forum Telfor (TELFOR), 2014 22nd*, pages 589–592, Nov 2014.

[KOG15]    K. Kralevska, H. Overby, and D. Gligoroski. Coded packet transport for optical packet/burst switched networks. In *2015 IEEE Global Communications Conference (GLOBECOM)*, pages 1–6, Dec 2015.

[KRH+08]   S. Katti, H. Rahul, W. Hu, D. Katabi, M. Medard, and J. Crowcroft. XORs in the air: Practical wireless network coding. *IEEE/ACM Trans. Netw*, 16(3):497–510, 2008.

[KRLL11]   A. E. Kamal, A. Ramamoorthy, L. Long, and S. Li. Overlay protection against link failures using network coding. *IEEE/ACM Trans. Netw*, 19(4):1071–1084, 2011.

[KSS11]   A. M. Kermarrec, N. Le Scouarnec, and G. Straub. Repairing multiple failures with coordinated and adaptive regenerating codes. In *International Symposium on Network Coding*, pages 1–6, July 2011.

[Law01]   E. Lawler. *Combinatorial Optimization : Networks and Matroids*. Dover Publications, 2001.

[LC83]   S. Lin and D. J. Costello. *Error Control Coding: Fundamentals and Applications*. Prentice-Hall, Englewood Cliffs, NJ, USA, 1983.

[LL14]   J. Li and B. Li. Cooperative repair with minimum-storage regenerating codes for distributed storage. In *IEEE Conference on Computer Communications (INFOCOM)*, pages 316–324, April 2014.

[LLL15]   R. Li, J. Lin, and P. P. C. Lee. Enabling concurrent failure recovery for regenerating-coding-based storage systems: From theory to practice. *IEEE Transactions on Computers*, 64(7):1898–1911, July 2015.

[LMS+97]   M. Luby, M. Mitzenmacher, M. A. Shokrollahi, D. A. Spielman, and V. Stemann. Practical loss-resilient codes. In *STOC*, pages 150–159. ACM, 1997.

[LMS09]   D. E. Lucani, M. Médard, and M. Stojanovic. Random linear network coding for time-division duplexing: Field size considerations. In *GLOBECOM*, pages 1–6, 2009.

[LMSS01a]   M. G. Luby, M. Mitzenmacher, M. A. Shokrollahi, and D. A. Spielman. Efficient erasure correcting codes. *IEEE Transactions on Information Theory*, 47(2):569–584, Feb 2001.

[LMSS01b]   M. G. Luby, M. Mitzenmacher, M. A. Shokrollahi, and D. A. Spielman. Improved low-density parity-check codes using irregular graphs. *IEEE Transactions on Information Theory*, 47(2):585–598, Feb 2001.

[LN86]   R. Lidl and H. Niederreiter. *Introduction to Finite Fields and Their Applications*. Cambridge University Press, New York, NY, USA, 1986.

[LPHF14a]   D. E. Lucani, M. V. Pedersen, J. Heide, and F. H. P. Fitzek. Coping with the upcoming heterogeneity in 5G communications and storage using fulcrum network codes. In *ISWCS*, pages 997–1001, 2014.

[LPHF14b]  D. E. Lucani, M. V. Pedersen, J. Heide, and F. H. P. Fitzek. Fulcrum network codes: A code for fluid allocation of complexity. *CoRR*, abs/1404.6620, 2014.

[LR10]  S. Li and A. Ramamoorthy. Improved compression of network coding vectors using erasure decoding and list decoding. *IEEE Communications Letters*, 14(8):749–751, 2010.

[Lub02]  M. Luby. LT codes. In *IEEE Symposium on Foundations of Computer Science (FOCS)*, 2002.

[LWLZ10]  Z. Liu, C. Wu, B. Li, and S. Zhao. UUSee: Large-scale operational on-demand streaming with random network coding. In *IEEE INFOCOM*, pages 2070–2078, 2010.

[LY82]  S. Lin and P.S. Yu. A hybrid arq scheme with parity retransmission for error control of satellite channels. *IEEE Transactions on Communications*, 30(7):1701–1719, July 1982.

[LYC03]  S. Y. R. Li, R. W. Yeung, and N. Cai. Linear network coding. *IEEE Transactions on Information Theory*, 49, 2003.

[MDXA10]  E. D. Manley, J.S. Deogun, L. Xu, and D. R. Alexander. All-optical network coding. *IEEE/OSA Journal of Optical Communications and Networking*, 2(4):175–191, April 2010.

[Men]  K. Menger. Zur allgemeinen kurventheorie.

[MFHH02]  S. Madden, M. J. Franklin, J. M. Hellerstein, and W. Hong. TAG: A tiny AGgregation service for ad-hoc sensor networks. In *OSDI*, pages 1–16, 2002.

[MG08]  R.C. Menendez and J.W. Gannet. Efficient, fault-tolerant all-optical multicast networks via network coding. In *Optical Fiber communication/National Fiber Optic Engineers Conference*, pages 1–3, Feb 2008.

[MN95]  D. J. C. MacKay and R. M. Neal. Good codes based on very sparse matrices. *Lecture Notes in Computer Science*, 1025, 1995.

[MN97]  D. J. C. MacKay and R. M. Neal. Near shannon limit performance of low density parity check codes. *Electronics Letters*, 33(6):457–458, Mar 1997.

[MNS14]  T. Miyamae, T. Nakao, and K. Shiozawa. Erasure code with shingled local parity groups for efficient recovery from multiple disk failures. In *10th Workshop on Hot Topics in System Dependability (HotDep)*. USENIX Association, 2014.

[MS78]  F.J. MacWilliams and N.J.A. Sloane. *The Theory of Error-Correcting Codes*. North-holland Publishing Company, 2nd edition, 1978.

[MS12]     M. Medard and A. Sprintson. *Network coding, Fundamentals and Applications*. 2012.

[NBG08]    M. Nord, S. Bjornstad, and C. M. Gauger. OPS or OBS in the core network?- A comparison of optical packet- and optical burst switching, 2008.

[NNC10]    K. Nguyen, T. P. Nguyen, and S.-C. S. Cheung. Video streaming with network coding. *Signal Processing Systems*, 59(3):319–333, 2010.

[NZ02]     T. Nguyen and A. Zakhor. Distributed video streaming with forward error correction, 2002.

[NZ03]     T. Nguyen and A. Zakhor. Path diversity with forward error correction (pdf) system for packet switched networks. In *INFOCOM*, volume 1, pages 663–672 vol.1, March 2003.

[OBBT12]   H. Overby, G. Biczok, P. Babarczi, and J. Tapolcai. Cost comparison of 1+1 path protection schemes: A case for coding. In *IEEE International Conference on Communications (ICC)*, 2012.

[OD11]     F. E. Oggier and A. Datta. Self-repairing homomorphic codes for distributed storage systems. In *INFOCOM*, pages 1215–1223, 2011.

[OLV$^+$12]  Paulo F. Oliveira, Luísa Lima, Tiago T. V. Vinhoza, João Barros, and Muriel Médard. Coding for trusted storage in untrusted networks. *IEEE Transactions on Information Forensics and Security*, 7(6):1890–1899, 2012.

[OSHT01]   M.J. O'Mahony, D. Simeonidou, D.K. Hunter, and A. Tzanakaki. The application of optical packet switching in future communication networks. *IEEE Communications Magazine*, 39(3):128–135, Mar 2001.

[Ove04]    H. Overby. Network layer packet redundancy in optical packet switched networks. *Opt. Express*, 12(20):4881–4895, Oct 2004.

[Ove08]    H. Overby. Combined study on survivability and performance in optical packet switched networks. *J. Opt. Netw.*, 7(4):294–309, Apr 2008.

[PB61]     W. W. Peterson and D. T. Brown. Cyclic codes for error detection. *Proceedings of the IRE*, 49(1):228–235, Jan 1961.

[PDC13]    D.S. Papailiopoulos, A.G. Dimakis, and V.R. Cadambe. Repair optimal erasure codes through hadamard designs. *IEEE Transactions on Information Theory*, 59(5):3021–3037, May 2013.

[PFS05]    P. Pakzad, C. Fragouli, and A. Shokrollahi. Coding schemes for line networks. In *Proc. ISIT*, 2005.

[PGK88]     D. A. Patterson, G. Gibson, and R. H. Katz. A case for redundant arrays of inexpensive disks (RAID). In *ACM SIGMOD International Conference on Management of Data*, pages 109–116, May 26–28 1988.

[PJBM$^+$16]     L. Pamies-Juarez, F. Blagojević, R. Mateescu, C. Gyuot, E. E. Gad, and Z. Bandić. Opening the chrysalis: On the real repair performance of msr codes. In *USENIX Conference on File and Storage Technologies (FAST)*, pages 81–94, February 2016.

[Pla05]     J. S. Plank. Erasure codes for storage applications. Tutorial Slides, presented at *FAST-2005: 4th Usenix Conference on File and Storage Technologies*, 2005.

[PLD$^+$12]     D.S. Papailiopoulos, J. Luo, A.G. Dimakis, C. Huang, and J. Li. Simple regenerating codes: Network coding for cloud storage. In *INFOCOM*, pages 2801–2805, March 2012.

[QY99]     C. Qiao and M. Yoo. Optical burst switching (obs) - a new paradigm for an optical internet. *Journal of High Speed Networks*, 8:69–84, 1999.

[Rii04]     S. Riis. Linear versus nonlinear boolean functions in network flow. *CISS*, 2004.

[RNW$^+$15]     K. V. Rashmi, Preetum Nakkiran, Jingyan Wang, Nihar B. Shah, and Kannan Ramchandran. Having your cake and eating it too: Jointly optimal erasure codes for I/O, storage, and network-bandwidth. In Jiri Schindler and Erez Zadok, editors, *Proceedings of the 13th USENIX Conference on File and Storage Technologies, FAST 2015, Santa Clara, CA, USA, February 16-19, 2015*, pages 81–94. USENIX Association, 2015.

[RS60]     I. S. Reed and G. Solomon. Polynomial codes over certain finite fields. *Journal of the Society for Industrial and Applied Mathematics*, 8(2):300–304, June 1960.

[RS02]     R. Ramaswami and K. N. Sivarajan. *Optical Networks: A Practical Perspective*. Second edition, 2002.

[RSG$^+$14]     K. V. Rashmi, N. B. Shah, D. Gu, H. Kuang, D. Borthakur, and K. Ramchandran. A "hitchhiker's" guide to fast and efficient data reconstruction in erasure-coded data centers. In *ACM SIGCOMM 2014 Conference, SIGCOMM'14, Chicago, IL, USA, August 17-22, 2014*, pages 331–342, 2014.

[RSK10]     K. V. Rashmi, N. B. Shah, and P. V. Kumar. Network coding. *Resonance*, 15(7):604–621, 2010.

[RSK11]     K. V. Rashmi, N. B. Shah, and P. V. Kumar. Optimal exact-regenerating codes for distributed storage at the msr and mbr points via a product-matrix construction. *IEEE Transactions on Information Theory*, 57(8):5227–5239, Aug 2011.

[RSR13]   K. V. Rashmi, N. B. Shah, and K. Ramchandran. A piggybacking design framework for read-and download-efficient distributed storage codes. *CoRR*, abs/1302.5872, 2013.

[RSU01]   T. J. Richardson, M. A. Shokrollahi, and R. L. Urbanke. Design of capacity-approaching irregular low-density parity-check codes. *IEEE Transactions on Information Theory*, 47(2):619–637, Feb 2001.

[SAK15]   B. Sasidharan, G. K. Agarwal, and P. V. Kumar. A high-rate MSR code with polynomial sub-packetization level. *CoRR*, abs/1501.06662, 2015.

[SAP+13]  M. Sathiamoorthy, M. Asteris, D. S. Papailiopoulos, A. G. Dimakis, R. Vadali, S. Chen, and D. Borthakur. XORing elephants: Novel erasure codes for big data. *PVLDB*, 6(5):325–336, 2013.

[Sho06]   A. Shokrollahi. Raptor codes. *IEEE Transactions on Information Theory*, 52(6):2551–2567, 2006.

[Sil12]   D. Silva. Minimum-overhead network coding in the short packet regime. In *International Symposium on Network Coding (NetCod)*, pages 173–178, June 2012.

[Sin64]   R. Singleton. Maximum distance q-nary codes. *IEEE Transactions on Information Theory*, 10(2):116–118, Apr 1964.

[SKFA09]  M. J. Siavoshani, L. Keller, C. Fragouli, and K. J. Argyraki. Compressed network coding vectors. In *ISIT*, pages 109–113, 2009.

[SKRC10]  K. Shvachko, H. Kuang, S. Radia, and R. Chansler. The hadoop distributed file system. In *IEEE Symposium on Mass Storage Systems and Technologies (MSST)*, pages 1–10, 2010.

[SMS12]   N. Sreenath, K. Muthuraj, and P. Sivasubramanian. Secure optical internet: Attack detection and prevention mechanism. In *International Conference on Computing, Electronics and Electrical Technologies (IC-CEET)*, pages 1009–1012, March 2012.

[SR10]    Changho Suh and Kannan Ramchandran. Exact regeneration codes for distributed storage repair using interference alignment. *CoRR*, abs/1001.0107, 2010.

[SRM02]   L. H. Sahasrabuddhe, S. Ramamurthy, and B. Mukherjee. Fault management in IP-over-WDM networks: WDM protection versus IP restoration. *IEEE Journal on Selected Areas in Communications*, 20(1):21–33, 2002.

[TCBOF11] O. Trullols-Cruces, J. M. Barcelo-Ordinas, and M. Fiore. Exact decoding probability under random linear network coding. *IEEE Communications Letters*, 15(1):67–69, January 2011.

[TF12]   N. Thomos and P. Frossard. Toward one symbol network coding vectors. *IEEE Communications Letters*, 16(11):1860–1863, 2012.

[TPD13]  I. Tamo, D. S. Papailiopoulos, and A. G. Dimakis. Optimal locally repairable codes and connections to matroid theory. *CoRR*, abs/1301.7693, 2013.

[Tur99]  J. S. Turner. Terabit burst switching. *J. High Speed Networks*, 8(1):3–16, 1999.

[TWB13]  I. Tamo, Z. Wang, and J. Bruck. Zigzag codes: MDS array codes with optimal rebuilding. *IEEE Transactions on Information Theory*, 59(3):1597–1616, 2013.

[TWB14]  I. Tamo, Zhiying Wang, and J. Bruck. Access versus bandwidth in codes for storage. *IEEE Transactions on Information Theory*, 60(4):2028–2037, April 2014.

[VB09]   F. Vieira and J. Barros. Network coding multicast in satellite networks. In *Next Generation Internet Networks*, pages 1–6, July 2009.

[VCR00]  S. Verma, H. Chaskar, and R. Ravikanth. Optical burst switching: a viable solution for terabit ip backbone. *IEEE Network*, 14(6):48–53, Nov 2000.

[VJS02]  V.M. Vokkarane, J.P. Jue, and S. Sitaraman. Burst segmentation: an approach for reducing packet loss in optical burst switched networks. In *IEEE International Conference on Communications (ICC)*, volume 5, pages 2673–2677, May 2002.

[VPFH10] P. Vingelmann, M. V. Pedersen, F. H. P. Fitzek, and J. Heide. Multimedia distribution using network coding on the iphone platform. In *ACM Multimedia Workshop on Mobile Cloud Media Computing*, MCMC '10, pages 3–6, 2010.

[VZ06]   V.M. Vokkarane and Q. Zhang. Forward redundancy: a loss recovery mechanism for optical burst-switched networks. In *International Conference on Wireless and Optical Communications Networks*, pages 5–10, 2006.

[WD09]   Y. Wu and A. G. Dimakis. Reducing repair traffic for erasure coding-based storage via interference alignment. In *ISIT*, pages 2276–2280, 2009.

[WK02]   H. Weatherspoon and J. Kubiatowicz. Erasure coding vs. replication: A quantitative comparison. In *International Workshop on Peer-to-Peer Systems (IPTPS), LNCS*, volume 1, 2002.

[WTB11]  Z. Wang, I. Tamo, and J. Bruck. On codes for optimal rebuilding access. In *Allerton Conference on Communication, Control, and Computing (Allerton)*, pages 1374–1381, Sept 2011.

[WWX10]   X. Wang, J. Wang, and Y. Xu. Data dissemination in wireless sensor networks with network coding. *EURASIP J. Wireless Comm. and Networking*, 2010.

[WXHO10]   X. Wang, Y. Xu, Y. Hu, and K. Ou. Mfr: Multi-loss flexible recovery in distributed storage systems. In *IEEE International Conference on Communications (ICC)*, pages 1–5, May 2010.

[YB16a]   M. Ye and A. Barg. Explicit constructions of high-rate MDS array codes with optimal repair bandwidth. *CoRR*, abs/1604.00454, 2016.

[YB16b]   M. Ye and A. Barg. Explicit constructions of optimal-access MDS codes with nearly optimal sub-packetization. *CoRR*, abs/1605.08630, 2016.

[YKSC01]   S.-W. Yuk, M.-G. Kang, B.-C. Shin, and D.-H. Cho. An adaptive redundancy control method for erasure-code-based real-time data transmission over the internet. *IEEE Transactions on Multimedia*, 3(3):366–374, Sep 2001.

[YSH$^+$98]   Y. Yamada, K. Sasayama, K. Habara, A. Misawa, M. Tsukada, T. Matsunaga, and K. Yukimatsu. Optical output buffered ATM switch prototype based on FRONTIERNET architecture. *IEEE Journal on Selected Areas in Communications*, 16(7):1298–1308, 1998.

[ZS00]   D. Zhou and S.S. Subramaniam. Survivability in optical networks. *IEEE Network*, 14(6):16–23, Nov 2000.

# Part II

# Included Papers

**Balanced XOR-ed Coding**
Katina Kralevska, Danilo Gligoroski, and Harald Øverby

**Families of Optimal Binary Non-MDS Erasure Codes**
Danilo Gligoroski and Katina Kralevska
IEEE Proceedings on International Symposium on Information Theory (ISIT), pp. 3150-3154, 2014

**Minimal Header Overhead for Random Linear Network Coding**

Danilo Gligoroski, Katina Kralevska, and Harald Øverby

IEEE International Conference on Communication Workshop (ICCW), pp. 680 - 685, 2015

**General Sub-packetized Access-Optimal Regenerating Codes**
Katina Kralevska, Danilo Gligoroski, and Harald Øverby

**HashTag Erasure Codes: From Theory to Practice**
Katina Kralevska, Danilo Gligoroski, Rune E.Jensen, and Harald Øverby
Submitted to IEEE Transactions on Big Data

**Balanced Locally Repairable Codes**
Katina Kralevska, Danilo Gligoroski, and Harald Øverby
International Symposium on Turbo Codes and Iterative Information Processing, 2016

**Coded Packet Transport for Optical Packet/Burst Switched Networks**

Katina Kralevska, Harald Øverby, and Danilo Gligoroski

IEEE Proceedings on Global Communications Conference (GLOBECOM), pp. 1 - 6, 2015